\documentclass[%
 reprint,
 amsmath,amssymb,
 aps,
]{revtex4-2}

\usepackage[T1]{fontenc}
\usepackage{graphicx}% Include figure files
\graphicspath{{/Users/09906318@nebraska.edu/Documents/Docu_Rab/UNL/Research/SkyrmLab/Publications/DMI-measurement/DMI_Paper/ImagesPlots}}
\usepackage{dcolumn}% Align table columns on decimal point
\usepackage{microtype}
\usepackage{bm}% bold math

\usepackage{float}
\newfloat{scheme}{htbp}{los}
\floatname{scheme}{Scheme}
\floatname{chart}{Chart}
\newfloat{graph}{htbp}{loh}
\newsavebox{\panelimg}
\usepackage{tikz}

\usepackage{chemformula} % Formulas using \ch{}

\usepackage[
    colorlinks=true, 
    linkcolor=blue,  % Color for equations, figures, and sections
    citecolor=blue,  % Color for bibliography citations
    urlcolor=blue    % Color for external URLs
]{hyperref}

\begin{document}

\preprint{APS/123-QED}

%%%%%%%%%%%%%%%%%%%%%%%%%%%%%%%%%%%%%%%%%%%%%%%%%%%%%%%%%%%%%%%%%%%%%
%% Author and title data
%%%%%%%%%%%%%%%%%%%%%%%%%%%%%%%%%%%%%%%%%%%%%%%%%%%%%%%%%%%%%%%%%%%%%
\title{Tuning the Dzyaloshinskii-Moriya interaction in CoFeB thin film heterostructures}

\author{Rabindra R Oliya}
\email{roliya2@huskers.unl.edu}
\author{Christopher Keck}
\author{Jing Guo}
\author{Robert Streubel}
\author{S. Adenwalla}
\affiliation{Department of Physics and Astronomy, University of Nebraska-Lincoln, Lincoln, Nebraska 68588, USA}

\author{Lia Compton}
\author{Kristen S Buchanan}
\affiliation{Department of Physics, Colorado State University, Colorado 80521, USA}

\date{\today}

\begin{abstract}
The Dzyaloshinskii-Moriya Interaction (DMI) is essential for stabilizing chiral spin textures in spintronics, yet achieving consistent quantitative measurements across different techniques remains a significant challenge. Heterostructures consisting of magnetic thin films sandwiched between non-magnetic heavy metals have been shown to possess substantial DMI; however, these vary in both magnitude and handedness for supposedly identical samples, indicating the sensitivity of DMI to both interfacial properties and measurement techniques. Here, systematic tuning of the DMI in ultra-thin magnetic heterostructures of Pt/CoFeB/Ta/Pt is achieved by varying the thickness of a tantalum sub-monolayer with thicknesses ranging from $0.1\,\mathrm{nm}$ to $0.4\,\mathrm{nm}$. The magnitude and sign of the DMI are measured using three independent methods: the parallel stripe domain model, edge nucleation field analysis, and Brillouin light scattering. The three independent approaches reveal excellent agreement within the experimental error. This ability to controllably tune and reliably quantify DMI provides a viable route toward engineering chiral magnetic textures, including domain walls and skyrmions, for next-generation spintronic devices. The necessity for precise measurements of magnetic parameters and careful attention to detail are discussed because calculations of the DMI for each technique depend on different parameters.
\end{abstract}


\maketitle

%%%%%%%%%%%%%%%%%%%%%%%%%%%%%%%%%%%%%%%%%%%%%%%%%%%%%%%%%%%%%%%%%%%%%
%% Main Text
%%%%%%%%%%%%%%%%%%%%%%%%%%%%%%%%%%%%%%%%%%%%%%%%%%%%%%%%%%%%%%%%%%%%%
\section{Introduction}

The Dzyaloshinskii--Moriya interaction (DMI) is an antisymmetric exchange mechanism arising from spin-orbit coupling (SOC) in magnetic systems with broken inversion symmetry~\cite{1Dzya1957,2Mori1960}. This interaction emerges both in noncentrosymmetric bulk B20 compounds, such as MnSi and FeGe, and at the interfaces of thin-film heterostructures where a ferromagnetic layer adjoins a heavy metal (HM) with strong SOC~\cite{3Fert1980,4Kuep2023,5Dien2020,6Roha2013}. The interaction is governed by the Hamiltonian $H_{\mathrm{DMI}} = \mathbf{d}_{ij} \cdot (\mathbf{S}_i \times \mathbf{S}_j)$, where $\mathbf{d}_{ij}$ is the Dzyaloshinskii--Moriya (DM) vector and $\mathbf{S}_i$ and $\mathbf{S}_j$ are neighboring spins. Consequently, the DMI favors orthogonal spin alignment and introduces a fixed chirality by energetically favoring either right handed or left handed rotation of the spin. This mechanism stabilizes topologically protected chiral spin textures including magnetic skyrmions and chiral domain walls (DWs), and plays a role in DW motion~\cite{7P1Guan2021}. As a result, it serves as a foundational building block for advanced spintronic devices~\cite{4Kuep2023,8Bogd1989,9Pere2014,10Magn2022}.

For practical device integration, thin-film heterostructures offer distinct advantages over bulk B20 compounds. In particular, interfacial engineering enables precise control over the DMI, magnetic anisotropy, and saturation magnetization~\cite{11Fran2014,12Legr2022,13Shen2019}. Furthermore, bulk B20 materials typically host skyrmions only within a narrow temperature and magnetic-field window. In contrast, thin-film systems can stabilize chiral spin textures at room temperature, making them far more viable for technological applications~\cite{14Ever2018,15More2016,16Woo2016}. Incorporating magnetically soft, amorphous alloys such as CoFeB into these multilayers further reduces the density of structural pinning sites and facilitates low-power DW motion~\cite{17Ojha2023,18Grub2022}. Ultimately, the microscopic origins of the DMI, its interfacial tunability, and its influence on magnetization dynamics continue to drive extensive research in condensed-matter physics and materials science~\cite{6Roha2013,9Pere2014,11Fran2014,14Ever2018}.

Stabilizing chiral magnetic textures in such heterostructures relies on a delicate balance among competing magnetic energy contributions: exchange interaction, effective perpendicular magnetic anisotropy, magnetostatic interactions, and the DMI. In ferromagnets lacking DMI, the DW surface energy density arises from the competition between exchange and anisotropy energies~\cite{19Hube1998} and is given by $\sigma_{\mathrm{DW}} = 4\sqrt{A_{\mathrm{ex}} K_{\mathrm{eff}}}$, where $A_{\mathrm{ex}}$ is the exchange stiffness and $K_{\mathrm{eff}}$ is the effective magnetic anisotropy~\cite{6Roha2013,14Ever2018,20Ross2006}. The presence of interfacial DMI modifies this expression to $\sigma_{\mathrm{DW}} = 4\sqrt{A_{\mathrm{ex}} K_{\mathrm{eff}}} - \pi D\cos(\psi)$, where $D$ is the effective DMI constant and $\psi$ is the internal DW magnetization angle~\cite{4Kuep2023}. The additional term favors a chiral configuration for magnetic textures, which have been experimentally demonstrated in ultrathin multilayers~\cite{15More2016} and comprehensively reviewed in the literature~\cite{15More2016, 21Fert2017}. The balance between the anistropy energy and the DMI defines a critical DMI energy constant $D_{\mathrm{c}} = \frac{4}{\pi}\sqrt{A_{\mathrm{ex}} K_{\mathrm{eff}}}$, allowing for  the stabilization of chiral structures either by increasing $D$ or minimizing $K_{eff}$~\cite{22Herv2018}.

The ability to  tune and quantify the DMI is essential for designing next-generation spintronic devices. Achieving an optimal DMI value requires precise control over the layer architecture, deposition conditions, and temperature~\cite{22Herv2018}. This optimization is further complicated by systematic discrepancies across experimental characterization methods. Specifically, techniques based on DW dynamics, spin-wave propagation, and spin-orbit torques frequently yield conflicting quantitative values for the DMI~\cite{4Kuep2023,10Magn2022,23Kuep2024}. Resolving these experimental discrepancies while reliably tuning the DMI remains a critical step in translating these materials into functional technology. In this work, we systematically tune the interfacial DMI by varying the stack architecture of heavy-metal/ferromagnet/heavy-metal (HM/FM/HM) heterostructures and cross-validate its magnitude using multiple characterization techniques. Specifically, we pair a magnetically soft, amorphous CoFeB layer, selected to minimize pinning, with asymmetric HM interfaces that provide strong SOC and tunable inversion asymmetry. Finally, we quantify and compare the resulting DMI strength using three complementary methods: Brillouin light scattering (BLS), edge-nucleation field analysis, and magnetic stripe-domain width analysis.

\section{Synthesis and Characterization}
Thin films heterostructures of  Cr($2\,\mathrm{nm}$)/\allowbreak Ta($1\,\mathrm{nm}$)/\allowbreak Pt($2\,\mathrm{nm}$)/\allowbreak Co$_{60}$Fe$_{20}$B$_{20}$($0.7\,\mathrm{nm}$)/\allowbreak Ta($t_{\mathrm{Ta}}$)/\allowbreak Pt($3\,\mathrm{nm}$) were deposited on $128^{\circ}$ Y-cut lithium niobate and silicon substrates using magnetron sputtering (see Fig.~\ref{fig:S1} of the supporting information). The HM layers provide strong SOC, while the structural inversion asymmetry is tuned by inserting a Ta dusting layer of varying thickness ($t_{\mathrm{Ta}} = 0.1$ to $0.4\,\mathrm{nm}$ in $0.1\,\mathrm{nm}$ steps) between the CoFeB and upper Pt layers. All four samples were deposited sequentially to minimize variations in film thickness and all measurements were performed on as-deposited (unannealed) samples. The magnetic properties and domain structures were characterized using vibrating sample magnetometry (VSM), full-field magneto-optical Kerr effect (MOKE) microscopy, and BLS.

\begin{figure}[htbp]
    \centering
    % --- TOP ROW ---
    \begin{minipage}[t]{0.48\linewidth}
        \begin{tikzpicture}
            \node[inner sep=0pt] (img) {\includegraphics[width=\linewidth]{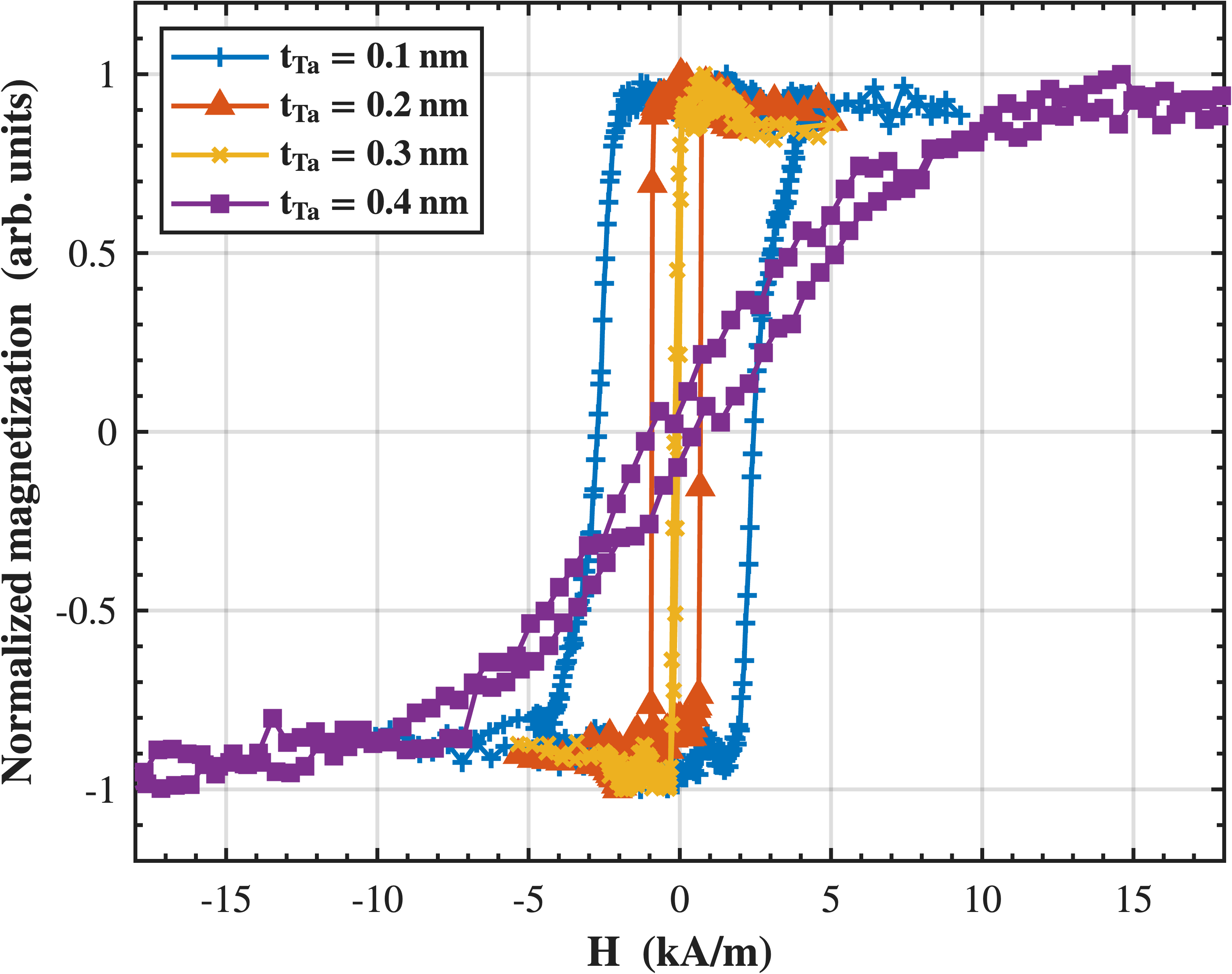}};
            \node[anchor=north west, xshift=-3mm, yshift=2mm, fill=none, fill opacity=0.5, text opacity=1, font=\small, inner sep=2pt] at (img.north west) {(a)};
        \end{tikzpicture}
    \end{minipage}\hfill
    \begin{minipage}[t]{0.48\linewidth}
        \begin{tikzpicture}
            \node[inner sep=0pt] (img) {\includegraphics[width=\linewidth]{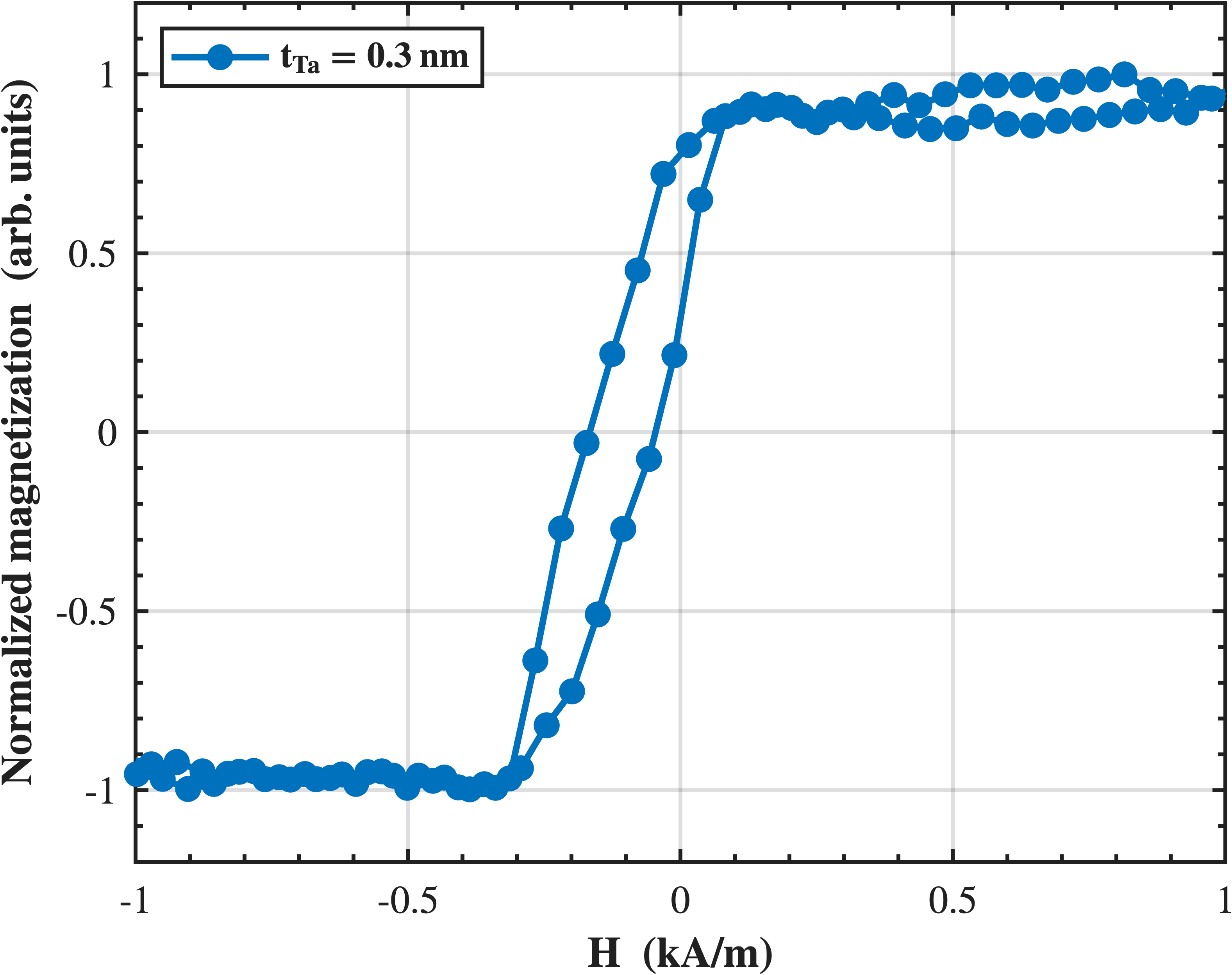}};
            \node[anchor=north west, xshift=-3mm, yshift=2mm, fill=none, fill opacity=0.5, text opacity=1, font=\small, inner sep=2pt] at (img.north west) {(b)};
        \end{tikzpicture}
    \end{minipage}

    \vspace{6mm} 

    % --- BOTTOM ROW ---
    \begin{minipage}[t]{0.48\linewidth}
        \begin{tikzpicture}
            \node[inner sep=0pt] (img) {\includegraphics[width=\linewidth]{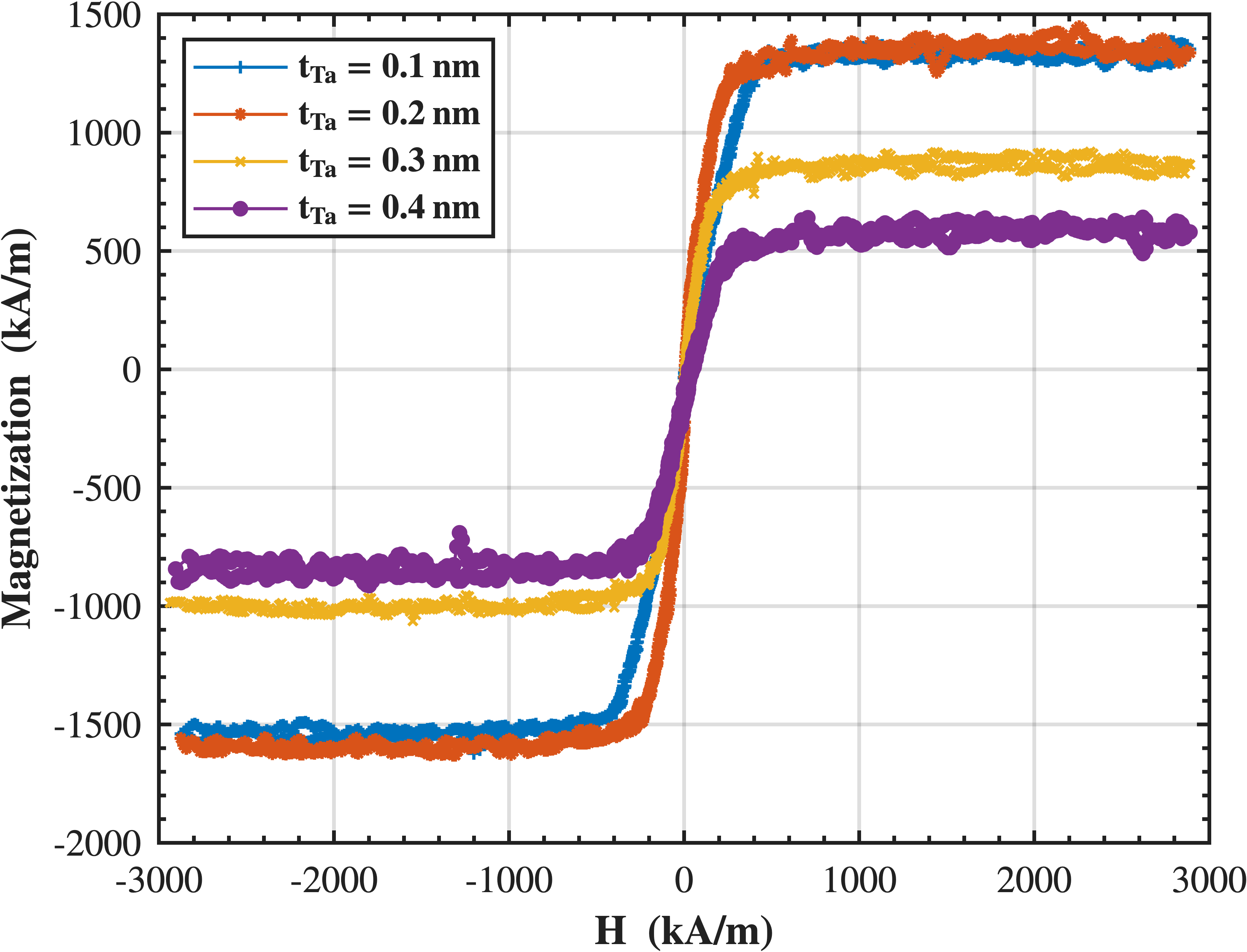}};
            \node[anchor=north west, xshift=-3mm, yshift=2mm, fill=none, fill opacity=0.5, text opacity=1, font=\small, inner sep=2pt] at (img.north west) {(c)};
        \end{tikzpicture}
    \end{minipage}\hfill
    \begin{minipage}[t]{0.48\linewidth}
        \begin{tikzpicture}
            \node[inner sep=0pt] (img) {\includegraphics[width=\linewidth]{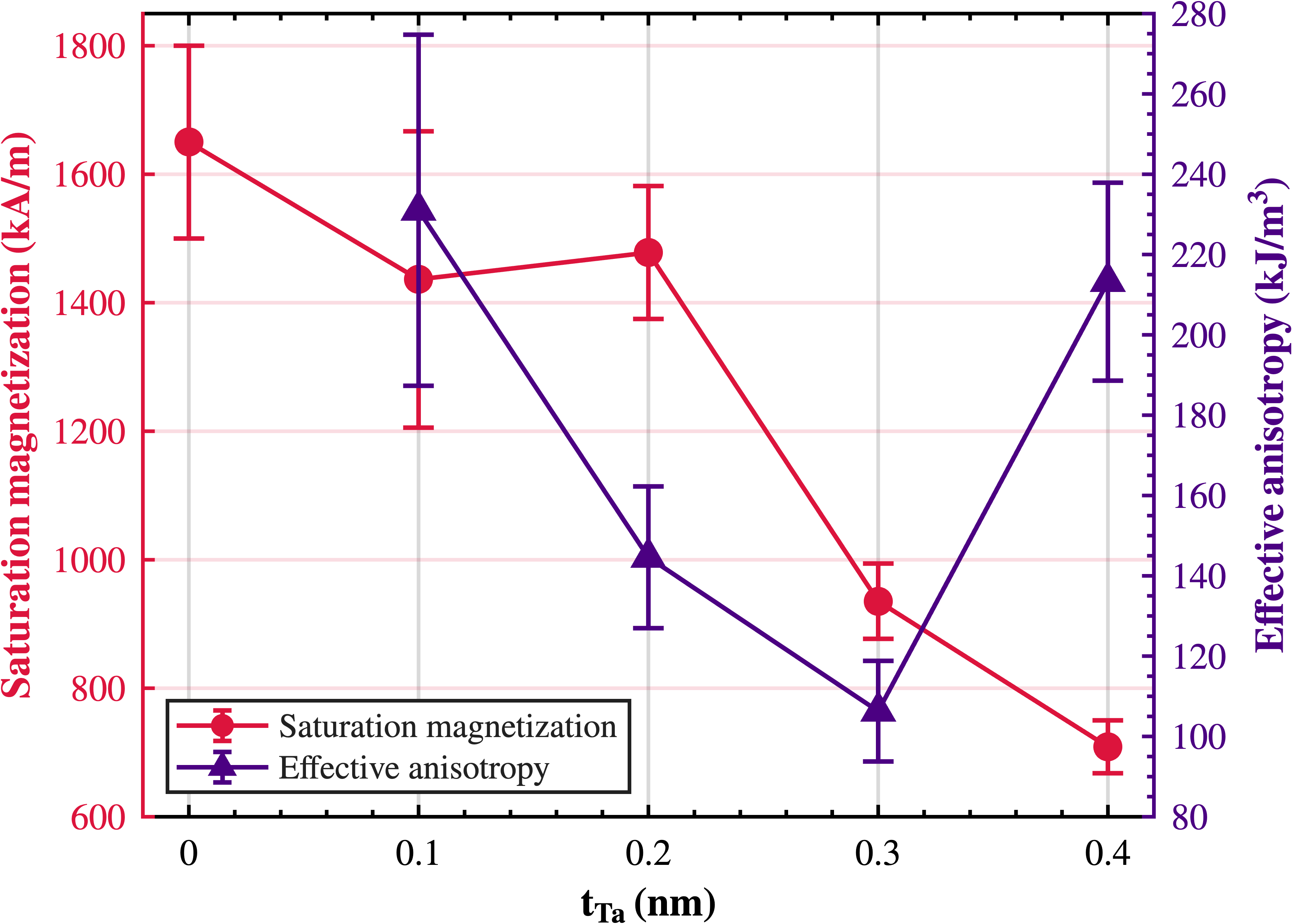}};
            \node[anchor=north west, xshift=-3mm, yshift=2mm, fill=none, fill opacity=0.5, text opacity=1, font=\small, inner sep=2pt] at (img.north west) {(d)};
        \end{tikzpicture}
    \end{minipage}

    \caption{Magnetic characterization of the CoFeB heterostructures as a function of Ta thickness ($t_{\mathrm{Ta}}$). 
    (a)~Normalized out-of-plane (OOP) magnetic hysteresis loops for samples with various Ta thicknesses, illustrating the modulation of coercivity and perpendicular magnetic anisotropy. 
    (b)~OOP hysteresis loop for $t_{\mathrm{Ta}} = 0.3\,\mathrm{nm}$. 
    (c)~Room-temperature ($300\,\mathrm{K}$) in-plane (IP) hysteresis loops measured via VSM for samples with different Ta thicknesses. 
    (d)~Measured saturation magnetization $M_{\mathrm{measured}}$ and effective perpendicular magnetic anisotropy $K_{\mathrm{eff}}$ as a function of $t_{\mathrm{Ta}}$.}
    \label{fig1}
\end{figure}

Figures~\ref{fig1}(a) and~\ref{fig1}(c) show the room temperature ($300\,\mathrm{K}$) out-of-plane (OOP) and in-plane (IP) hysteresis loops for samples with varying Ta thickness. The thickness of the Ta dusting layer strongly affects both the net saturation magnetization and the magnetic anisotropy. Prior experiments have shown that diffusion and alloying at the Ta/CoFeB interface form a magnetic dead layer~\cite{13Shen2019,24Tani2017}. This effect is evident in the monotonic decrease in sample magnetization with increasing Ta thickness. Assuming a constant intrinsic saturation magnetization ($M_{\mathrm{s}}$), the effective magnetic thickness of the CoFeB layer (see Fig.~\ref{fig:S2}) is estimated as $t_{\mathrm{eff}} = (M_{\mathrm{measured}}/M_{\mathrm{ref}}) \times t$. Here, $M_{\mathrm{measured}}$ is the measured saturation magnetization and $M_{\mathrm{ref}} = 1650\,\mathrm{kA/m}$~\cite{25Lavr2009} is the saturation magnetization of a control Pt/CoFeB/Pt thin film. All DMI values are evaluated using $M_{\mathrm{ref}}$ and $t_{\mathrm{eff}}$. The effective perpendicular magnetic anisotropy $K_{\mathrm{eff}}$ is determined by evaluating the area between the hard and easy axis hysteresis loops, as described previously~\cite{26Blan2005}, and is plotted as a function of Ta thickness in Fig.~\ref{fig1}(d). 

We determined the DMI of the samples using three complementary methods: edge-nucleation field analysis, stripe-domain width analysis, and BLS. Because the domain based techniques are constrained by domain size, the first two methods were restricted to the single sample with a $0.3\,\mathrm{nm}$ Ta dusting layer. In contrast, BLS measurements were performed on all four samples. Kerr microscopy enables direct visualization of the magnetic domain structures and is used for both the edge-nucleation and domain-width determinations of the DMI. Magnetic domain imaging was performed using a modified Zeiss Axiotron microscope and a full-field Kerr microscope (Evico Magnetics GmbH). These instruments can apply an OOP field up to $\pm 80.00\,\mathrm{kA/m}$ and an IP field up to $\pm 55.00\,\mathrm{kA/m}$, enabling controlled observation of domain evolution. Representative MOKE images of the samples are shown in Figs.~\ref{fig:S3}--\ref{fig:S5} (see supporting information). All MOKE images shown here were obtained after subtracting a reference image recorded at magnetic saturation.

\section{DMI Measurement}
\subsection{Edge Nucleation}
\begin{figure}[htbp]
    \centering
    % --- TOP ROW ---
    \begin{minipage}[t]{0.48\linewidth}
        \centering
        {\tiny $H_x = -36.76 \mathrm{kA/m},\, H_z = 1.64 \mathrm{kA/m}$}\par\vspace{1mm}
        \begin{tikzpicture}
            \node[inner sep=0pt, draw=black, thick] (img) {\includegraphics[width=\linewidth]{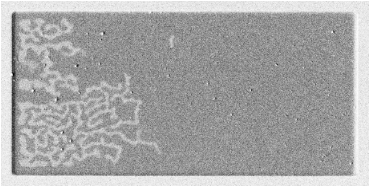}};
            \node[anchor=north west, fill=white, fill opacity=0.5, text opacity=1, font=\footnotesize, inner sep=2pt] at (img.north west) {(a)};
            \node[anchor=north east, fill=white, fill opacity=0.6, text opacity=1, font=\large, inner sep=1.5pt] at ([xshift=-2mm, yshift=-8mm]img.north east) {$\Leftarrow\,\odot$};
        \end{tikzpicture}
    \end{minipage}\hfill
    \begin{minipage}[t]{0.48\linewidth} 
        \centering
        {\tiny $H_x = 37.41 \mathrm{kA/m},\, H_z = 2.63 \mathrm{kA/m}$}\par\vspace{1mm}
        \begin{tikzpicture}
            \node[inner sep=0pt, draw=black, thick] (img) {\includegraphics[width=\linewidth]{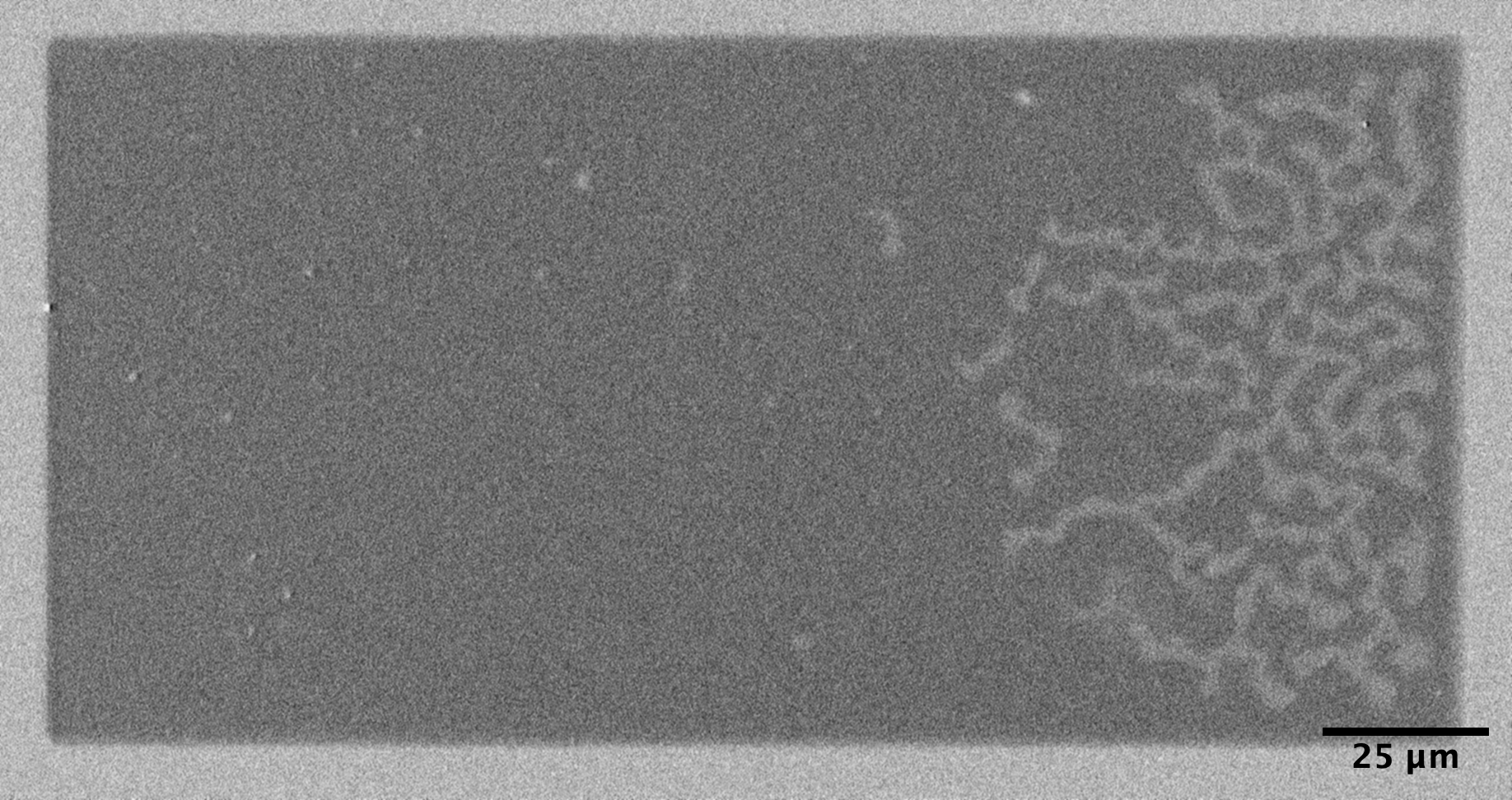}};
            \node[anchor=north west, fill=white, fill opacity=0.5, text opacity=1, font=\footnotesize, inner sep=2pt] at (img.north west) {(b)};
            \node[anchor=north west, fill=white, fill opacity=0.6, text opacity=1, font=\large, inner sep=1.5pt] at ([xshift=2mm, yshift=-8mm]img.north west) {$\odot\,\Rightarrow$};
        \end{tikzpicture}
    \end{minipage}

    \vspace{4mm} 

    % --- MIDDLE ROW ---
    \begin{minipage}[t]{0.48\linewidth}
        \centering
        {\tiny $H_x = -36.76 \mathrm{kA/m},\, H_z = -2.01 \mathrm{kA/m}$}\par\vspace{1mm}
        \begin{tikzpicture}
            \node[inner sep=0pt, draw=black, thick] (img) {\includegraphics[width=\linewidth]{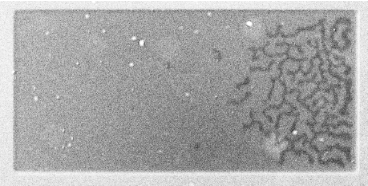}};
            \node[anchor=north west, fill=white, fill opacity=0.5, text opacity=1, font=\footnotesize, inner sep=2pt] at (img.north west) {(c)};
            \node[anchor=south west, fill=white, fill opacity=0.6, text opacity=1, font=\large, inner sep=1.5pt] at ([xshift=2mm, yshift=2mm]img.south west) {$\Leftarrow\,\otimes$};
        \end{tikzpicture}
    \end{minipage}\hfill
    \begin{minipage}[t]{0.48\linewidth} 
        \centering
        {\tiny $H_x = 37.41 \mathrm{kA/m},\, H_z = -1.04 \mathrm{kA/m}$}\par\vspace{1mm}
        \begin{tikzpicture}
            \node[inner sep=0pt, draw=black, thick] (img) {\includegraphics[width=\linewidth]{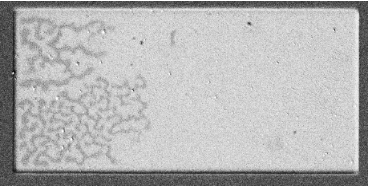}};
            \node[anchor=north west, fill=white, fill opacity=0.5, text opacity=1, font=\footnotesize, inner sep=2pt] at (img.north west) {(d)};
            \node[anchor=south east, fill=white, fill opacity=0.6, text opacity=1, font=\large, inner sep=1.5pt] at ([xshift=-2mm, yshift=2mm]img.south east) {$\otimes\,\Rightarrow$};
        \end{tikzpicture}
    \end{minipage}

    \vspace{6mm}

    % --- BOTTOM ROW ---
    \begin{minipage}[c]{0.22\linewidth}
        \centering
        \begin{tikzpicture}
            \node[inner sep=0pt] (img) {\includegraphics[width=\linewidth]{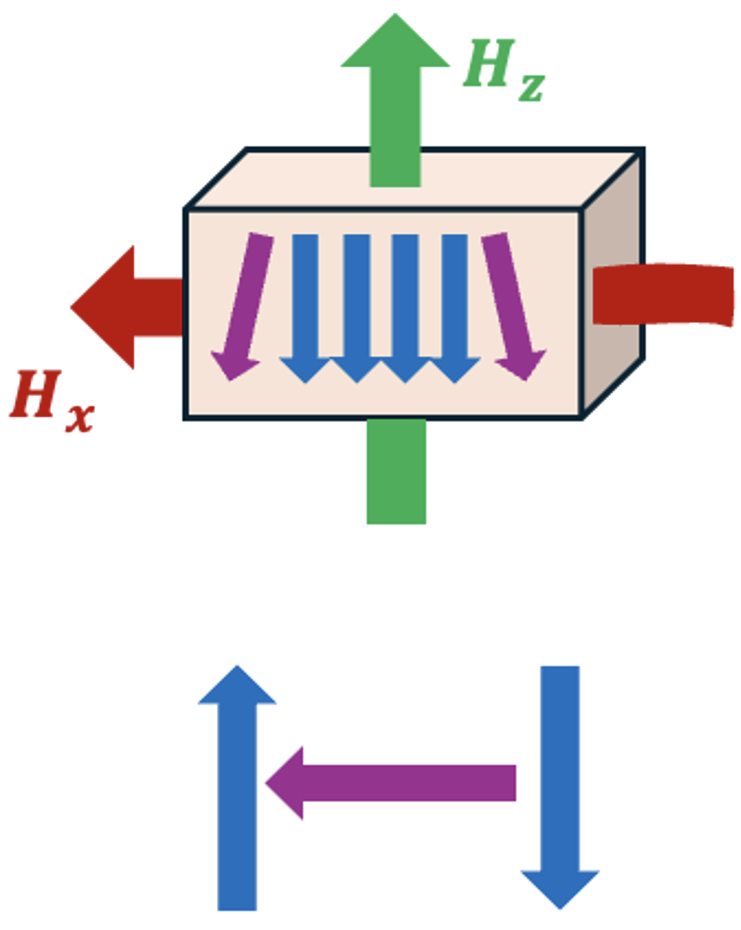}};
            \node[anchor=north west, fill=none, fill opacity=0.5, text opacity=1, font=\footnotesize, inner sep=1pt] at (img.north west) {(e)};
        \end{tikzpicture}
        
        \vspace{4mm}
        
        \begin{tikzpicture}
            \node[inner sep=0pt] (img) {\includegraphics[width=\linewidth]{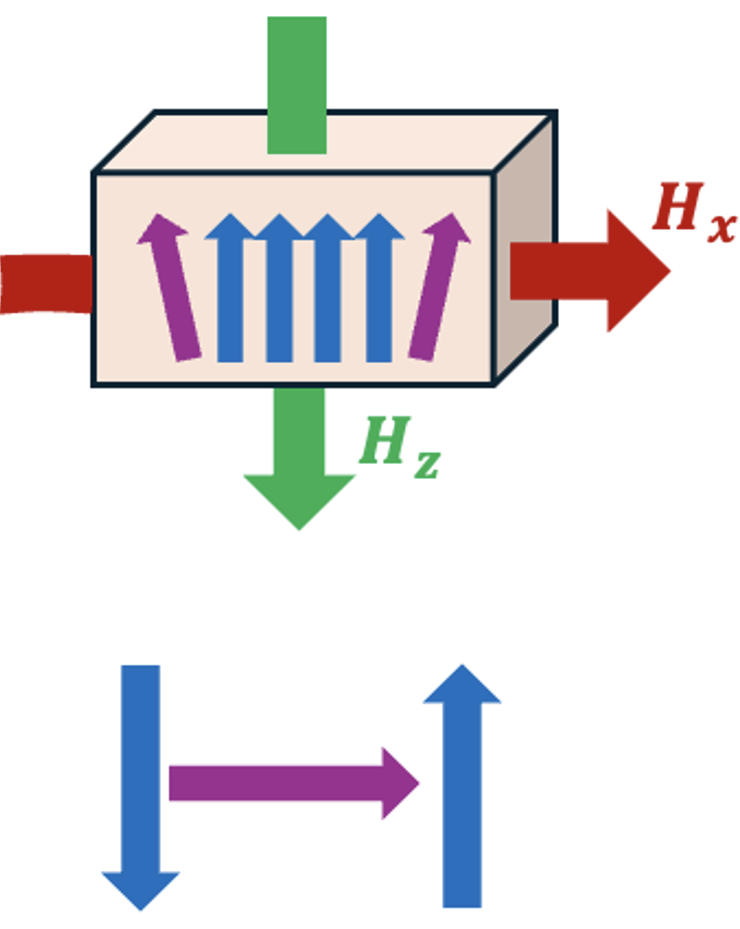}};
            \node[anchor=north west, fill=none, fill opacity=0.5, text opacity=1, font=\footnotesize, inner sep=1pt] at (img.north west) {(f)};
        \end{tikzpicture}
    \end{minipage}\hfill
    \begin{minipage}[c]{0.74\linewidth}
        \centering
        \begin{tikzpicture}
            \node[inner sep=0pt] (img) {\includegraphics[width=\linewidth]{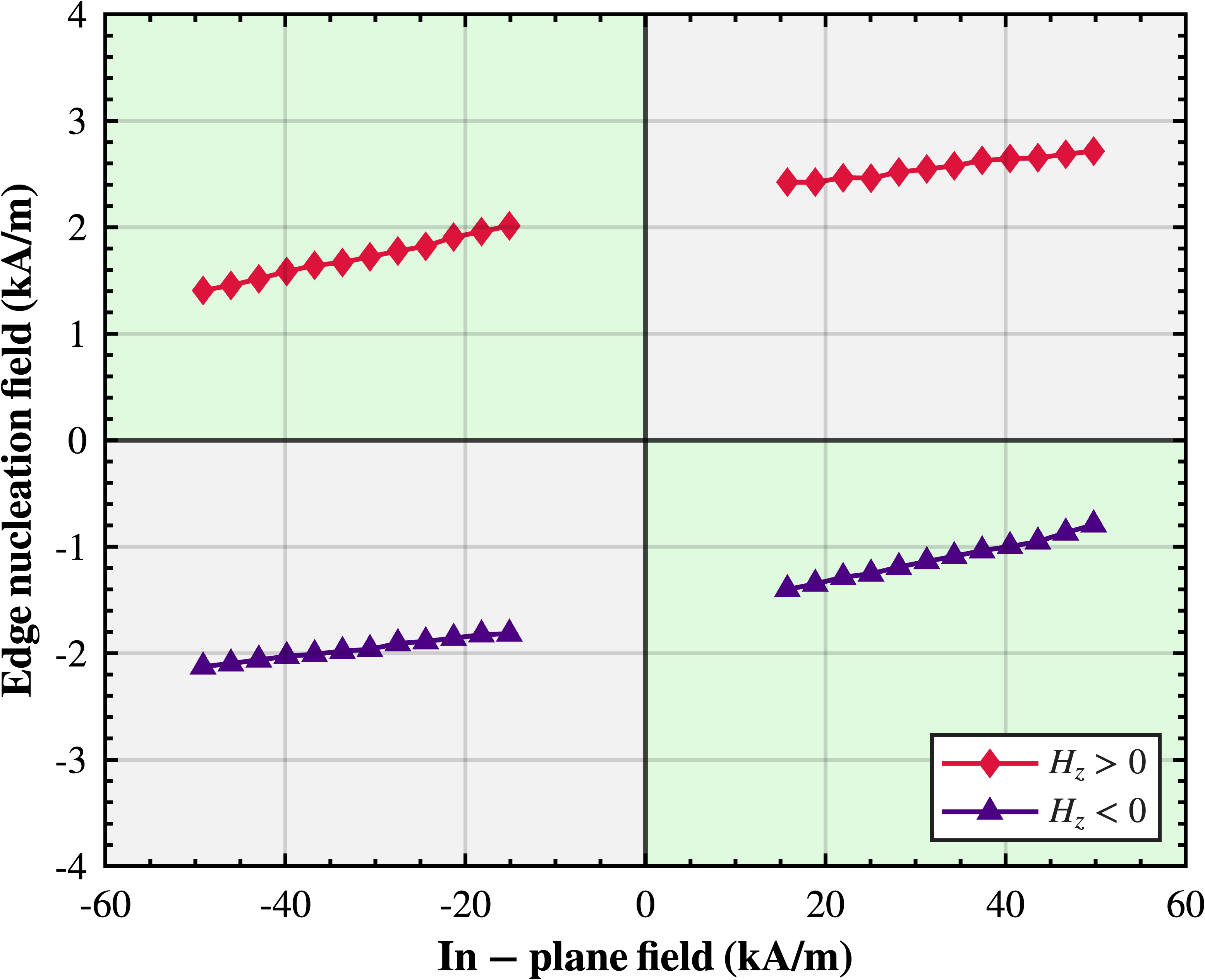}};
            \node[anchor=north west, xshift = -3mm, yshift= 2mm, fill=none, fill opacity=0.5, text opacity=1, font=\footnotesize, inner sep=2pt] at (img.north west) {(g)};
            
            % --- QUADRANT NUMERICAL LABELS ---
            \node[fill=white, fill opacity=0.6, text opacity=1, font=\scriptsize, inner sep=1pt] at ([xshift=6mm, yshift=-4mm]img.north) {(I)};
            \node[fill=white, fill opacity=0.6, text opacity=1, font=\scriptsize, inner sep=1pt] at ([xshift=-2mm, yshift=-4mm]img.north) {(II)};
            \node[fill=white, fill opacity=0.6, text opacity=1, font=\scriptsize, inner sep=1pt] at ([xshift=-2mm, yshift=10mm]img.south) {(III)};
            \node[fill=white, fill opacity=0.6, text opacity=1, font=\scriptsize, inner sep=1pt] at ([xshift=6mm, yshift=10mm]img.south) {(IV)};
            
            % --- NUCLEATION TEXT LABELS ---
            \node[rotate=5, fill=none, fill opacity=0.4, text opacity=1, font=\tiny, inner sep=1pt] at ([xshift=18mm, yshift=14mm]img.center) {Right edge nucleation};
            \node[rotate=14, fill=none, fill opacity=0.4, text opacity=1, font=\tiny, inner sep=1pt] at ([xshift=-12mm, yshift=10mm]img.center) {Left edge nucleation};
            \node[rotate=5, fill=none, fill opacity=0.4, text opacity=1, font=\tiny, inner sep=1pt] at ([xshift=-12mm, yshift=-11mm]img.center) {Right edge nucleation};
            \node[rotate=10, fill=none, fill opacity=0.4, text opacity=1, font=\tiny, inner sep=1pt] at ([xshift=18mm, yshift=-6mm]img.center) {Left edge nucleation};
            
        \end{tikzpicture}
    \end{minipage}

    \vspace{2mm} 

    \caption{Edge nucleation measurements under an IP field reveal the chiral nature of domain formation influenced by the DMI. Up domain (a) \& (b) nucleation occurs on the side of the rectangle aligned with the in-plane field direction, while down domains (c) \& (d) nucleate from the opposite side, confirming the intrinsic chirality of the system. 
        (e)--(f) Schematic illustration of domain nucleation and wall alignment in a DMI system under combined OOP and IP fields. (e) For a negatively saturated sample with OOP field upward and IP field leftward, nucleation begins at the left edge. (f) For a positively saturated sample with OOP field downward and IP field rightward, nucleation starts at the left edge. In both cases, domain walls align parallel to the IP field, indicating left-handed chirality and a negative DMI. 
        (g) Edge nucleation field as a function of applied IP field. Quadrants I and II represent the nucleation of up domains, and quadrants III and IV represent the nucleation of down domains.
    }
    
    \label{fig2}
\end{figure}

The edge-nucleation method~\cite{27Pizz2014,28Adhi2025}, like the bubble-domain expansion method~\cite{29Souc2016,30Khan2016}, relies on the chirality of domains formed in the presence of DMI and their nucleation behavior under an applied IP field. When both DMI and an IP field are present, domains nucleate preferentially along one edge of a lithographically defined structure. As illustrated schematically in Figs.~\ref{fig2}(e) and~\ref{fig2}(f), the DMI causes the magnetization at the edges of a magnetically saturated sample to tilt slightly, rotating either clockwise or counterclockwise according to the chirality. Applying an IP field then lowers the energy barrier for nucleation at one specific edge. Consequently, the sign of the DMI can be directly determined by observing domain nucleation under combined OOP and IP fields and the magnitude can be extracted by mapping the OOP and IP fields at which domains nucleate.

\begin{figure}[htbp]
    \centering
    % --- Top Row: Images (a) and (b) side by side ---
    \begin{minipage}[b]{0.32\linewidth}
        \centering
        \begin{tikzpicture}
            \node[anchor=south west, inner sep=0] (imgA) at (0,0) {\includegraphics[width=\linewidth]{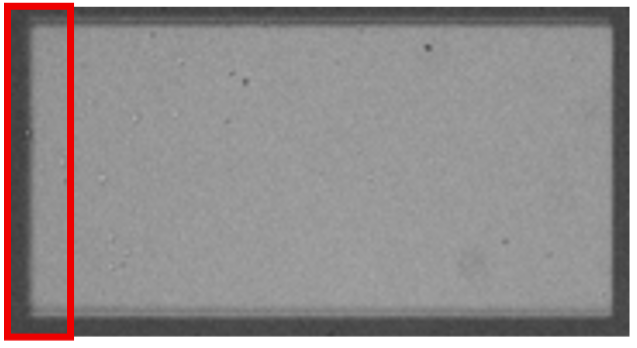}};
            \node[anchor=north west, fill=white, fill opacity=0.5, text opacity=1, inner sep=2pt, xshift=-1mm, yshift=4mm] at (imgA.north west) {\small (a)};
        \end{tikzpicture}
    \end{minipage}\hfill
    \begin{minipage}[b]{0.64\linewidth}
        \centering
        \begin{tikzpicture}
            \node[anchor=south west, inner sep=0] (imgB) at (0,0) {\includegraphics[width=\linewidth, height=2.0cm, keepaspectratio]{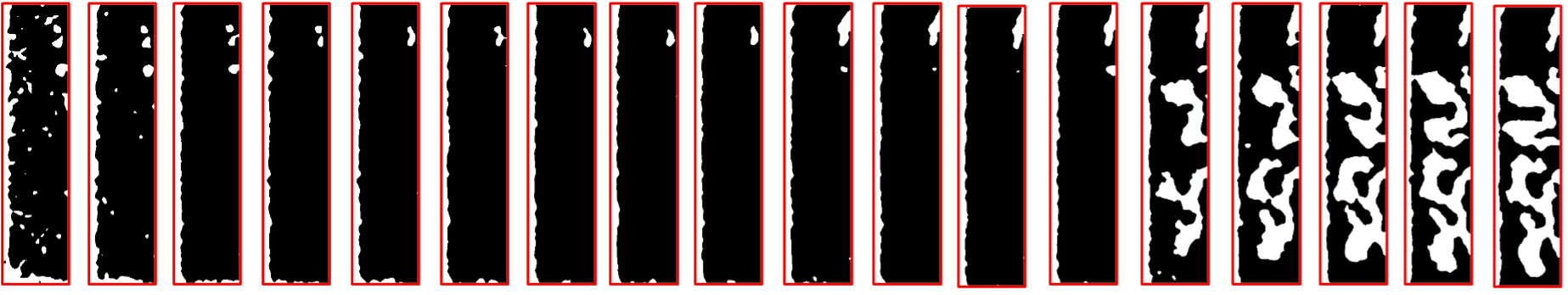}};
            \node[anchor=north west, fill=white, fill opacity=0.5, text opacity=1, inner sep=2pt, xshift=-1mm, yshift=4mm] at (imgB.north west) {\small (b)};
        \end{tikzpicture}
    \end{minipage}
    
    \vspace{4mm} 
    
    % --- Bottom Row: Graph (c) with insets ---
    \begin{minipage}[b]{\linewidth}
        \centering
        \begin{tikzpicture}
            \node[anchor=south west, inner sep=0] (graph) at (0,0) {\includegraphics[width=\linewidth]{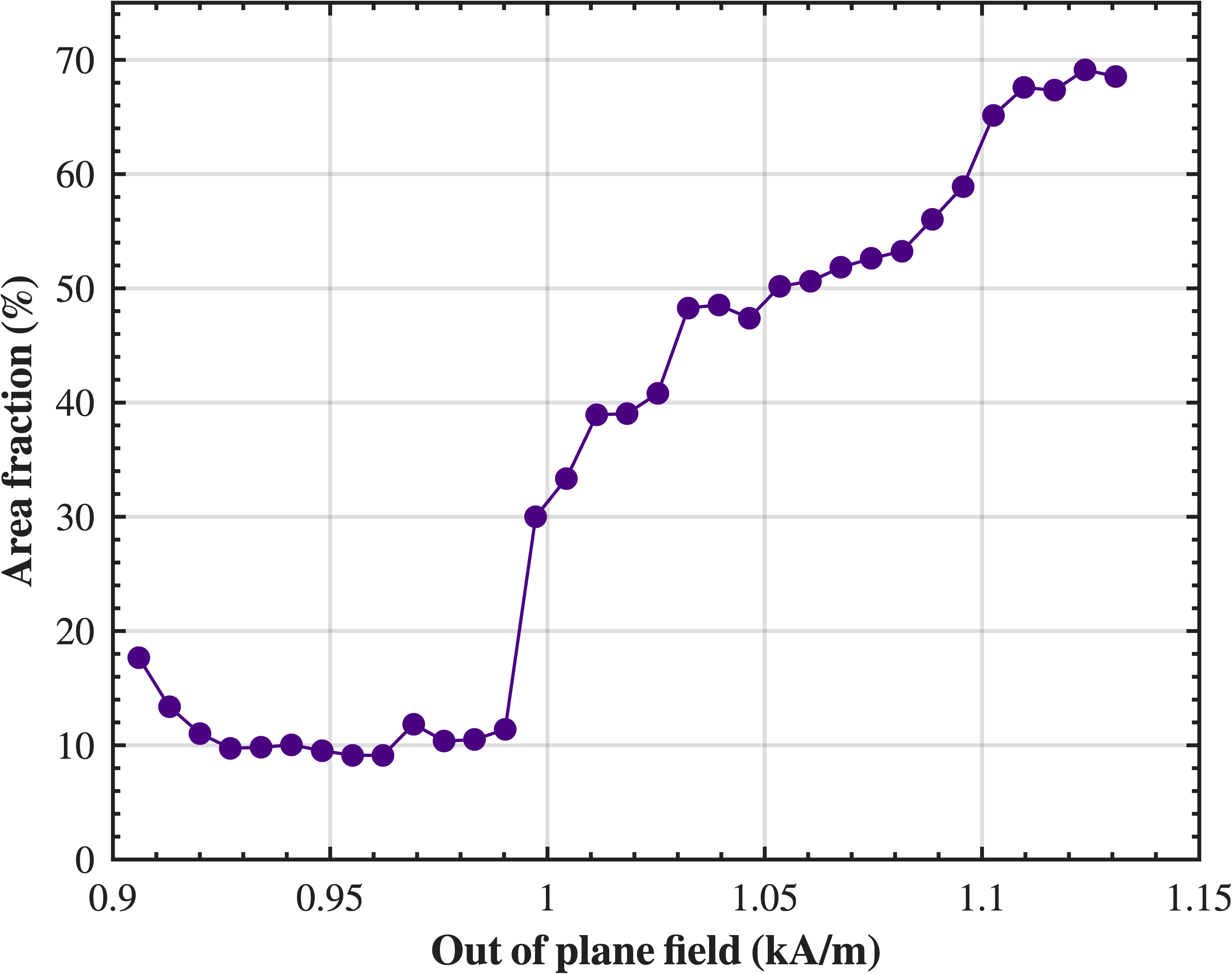}};
            \useasboundingbox (0,0) rectangle (graph.north east);
            
            % LABEL (c)
            \node[anchor=north west, fill=white, fill opacity=0.5, text opacity=1, inner sep=2pt, xshift=-1mm, yshift=2mm] at (graph.north west) {\small (c)};

            \begin{scope}[x={(graph.south east)}, y={(graph.north west)}]
                
                % Inset 1
                \node[inner sep=0, draw=red, line width=1.2pt] (inset1) at (0.22, 0.62) {\includegraphics[height=1.8cm]{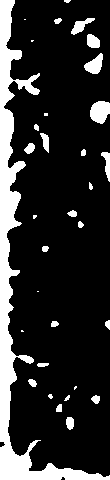}};
                \draw[-stealth, magenta, line width=1.5pt] (inset1.south) -- (0.12, 0.35);

                % Inset 2
                \node[inner sep=0, draw=red, line width=1.2pt] (inset2) at (0.35, 0.62) {\includegraphics[height=1.8cm]{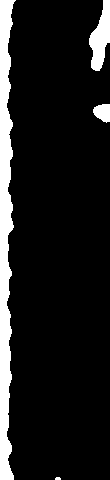}};
                \draw[-stealth, magenta, line width=1.5pt] (inset2.south) -- (0.40, 0.32);

                % Inset 3
                \node[inner sep=0, draw=red, line width=1.2pt] (inset3) at (0.80, 0.62) {\includegraphics[height=1.8cm]{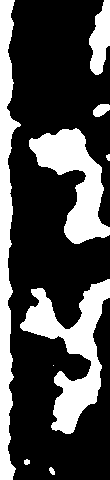}};
                \draw[-stealth, magenta, line width=1.5pt] (inset3.west) -- (0.45, 0.45);
                
            \end{scope}
        \end{tikzpicture}
    \end{minipage}
    
    \vspace{2mm} 
    
    \caption{Determination of the edge-nucleation field using MOKE microscopy. 
    (a)~MOKE image of the patterned rectangular structure. Regions of interest (ROIs) are defined along both edges to quantify the local nucleation field, with a representative ROI indicated by the red box on the left edge. 
    (b)~Binarized image of the magnetic domains captured while pulsing the OOP field with increasing amplitude under a fixed IP field. 
    (c)~Domain area fraction within the ROI as a function of the applied OOP field. The nucleation field is extracted from this curve and defined as the threshold OOP field at which a domain first appears within the ROI under a constant IP field. The three insets display the magnetic states corresponding to the specific OOP fields indicated by the arrows.}
    \label{fig3}
\end{figure}

To establish a well-defined nucleation boundary, the sample was lithographically patterned into a $200\,\mu\mathrm{m} \times 100\,\mu\mathrm{m}$ rectangle. In each measurement, the sample is first saturated in an OOP field, after which a constant IP field is applied. The opposing OOP field is then pulsed progressively with increasing amplitude. The nucleation field is defined as the threshold OOP field at which a domain first appears at an edge under a constant IP field (see Fig.~\ref{fig3}). Notably, the direction of the IP field plays a crucial role in controlling the location of domain nucleation.

Figures~\ref{fig2}(a) and~\ref{fig2}(b) illustrate the nucleation of up domains. When the IP field points to the left, up-domain nucleation initiates along the left edge of the rectangular structure. Conversely, a rightward IP field drives nucleation from the right edge. In contrast, down domains exhibit the opposite behavior, nucleating along the edge opposite to the applied IP field direction, as shown in Figs.~\ref{fig2}(c) and~\ref{fig2}(d). Furthermore, rotating the sample by $90^{\circ}$ relative to the IP field preserves this spatial asymmetry, with nucleation shifting to the long edge of the rectangle. This observation confirms that the asymmetric nucleation is an intrinsic physical phenomenon rather than an artifact of structural edge defects or magnetic-field tilt.

The schematics in Figs.~\ref{fig2}(e) and~\ref{fig2}(f) represent the zero-field edge magnetization configurations that are consistent with our observations. In Fig.~\ref{fig2}(e), the sample is initially negatively saturated. Applying an upward OOP field and a leftward IP field triggers magnetization reversal at the left edge, with the DW magnetization aligned parallel to the IP field. An analogous process occurs when the sample is initially positively saturated, as shown in Fig.~\ref{fig2}(f). Under a downward OOP field and a rightward IP field, nucleation again begins at the left edge of the sample. Based on these observations and the sign convention adopted from Ref.~\cite{4Kuep2023}, the DMI is negative and the domains exhibit left-handed chirality.

The dependence of the edge-nucleation field on the applied IP field is plotted in Fig.~\ref{fig2}(g). Nucleation initiates from the left edge in the second and fourth quadrants and from the right edge in the first and third quadrants. The nucleation field magnitudes are extracted through a consistent quantitative analysis of the minority-domain fractional area, as shown in Fig.~\ref{fig3}. The sharp onset of nucleation provides a clear threshold, enabling an accurate determination of the nucleation field across the $H_{\mathrm{IP}}$--$H_{\mathrm{OOP}}$ plane. The magnitude of the DMI is determined by fitting the data in the second and fourth quadrants to the edge-nucleation model~\cite{27Pizz2014,28Adhi2025}:
\begin{equation}
\resizebox{\linewidth}{!}{%
  $\displaystyle \mu_0 H_{n,\,\mathrm{edge}} = \frac{2\pi A_{\mathrm{ex}} t}{p k_{\mathrm{B}} T} \left(1 - \frac{D}{D_{\mathrm{c}}}\right)^2 \mu_0 H_k \left(1 \mp \frac{\mu_0 H_x}{\mu_0 H_k \left(1 - \frac{D}{D_{\mathrm{c}}}\right)}\right)^2$%
}
\label{equation1}
\end{equation}
\noindent where $H_k = 2K_{\mathrm{eff}}/\mu_0 M_{\mathrm{ref}}$. These two quadrants correspond to easy-edge nucleation, where the nucleation field decreases with increasing IP field magnitude. In the remaining two quadrants, domains that nucleate preferentially at the easy edge under lower fields propagate rapidly across the structure. As a result, they completely reverse the magnetization of the film before the threshold field for hard edge nucleation is reached. In our fits, we use the literature exchange stiffness $A_{\mathrm{ex}} = 27.5\,\mathrm{pJ/m}$~\cite{31Bhat2023}, $M_{\mathrm{ref}} = 1650\,\mathrm{kA/m}$~\cite{25Lavr2009}, and $K_{\mathrm{eff}} = 106.28\,\mathrm{kJ/m^3}$ [obtained from VSM measurements; see Fig.~\ref{fig1}(d)]. Using these parameters, we obtain a DMI magnitude of $1.86 \pm 0.22\,\mathrm{mJ/m^2}$.

\subsection{Parallel-Stripe Model for DMI Measurements}
In the ultrathin-film limit at zero field, the total energy of a parallel stripe-domain pattern is minimized for a half-periodicity of $L_0 = d \frac{\pi}{2} \exp\left[\frac{\sigma_{\mathrm{DW}}}{4\Lambda t} - \frac{1}{2}\right]$. Here, $t$ is the magnetic film thickness, $\sigma_{\mathrm{DW}}$ is the DW energy, and $\Lambda = (\mu_0/4\pi)M_{\mathrm{s}}^2$ is the dipolar energy constant~\cite{32Sara2010}. A sizable DMI alters the domain width by modifying the DW energy density according to $\sigma_{\mathrm{DW}} = 4\sqrt{A_{\mathrm{ex}}K_{\mathrm{eff}}} - \pi D \cos(\psi)$~\cite{4Kuep2023,6Roha2013}. We extract the half-periodicity of the domain pattern using a two-dimensional fast Fourier transform (FFT) of the MOKE domain images (see Fig.~\ref{fig:S6} of the supporting information). To rule out the effects of local sample inhomogeneities, two distinct regions of the MOKE image are analyzed.

\begin{figure}[htbp]
    \centering
    % --- TOP ROW: Figure (a) spanning full width ---
    \begin{tikzpicture}
        \node[inner sep=0pt] (img) {\includegraphics[width=0.75\linewidth]{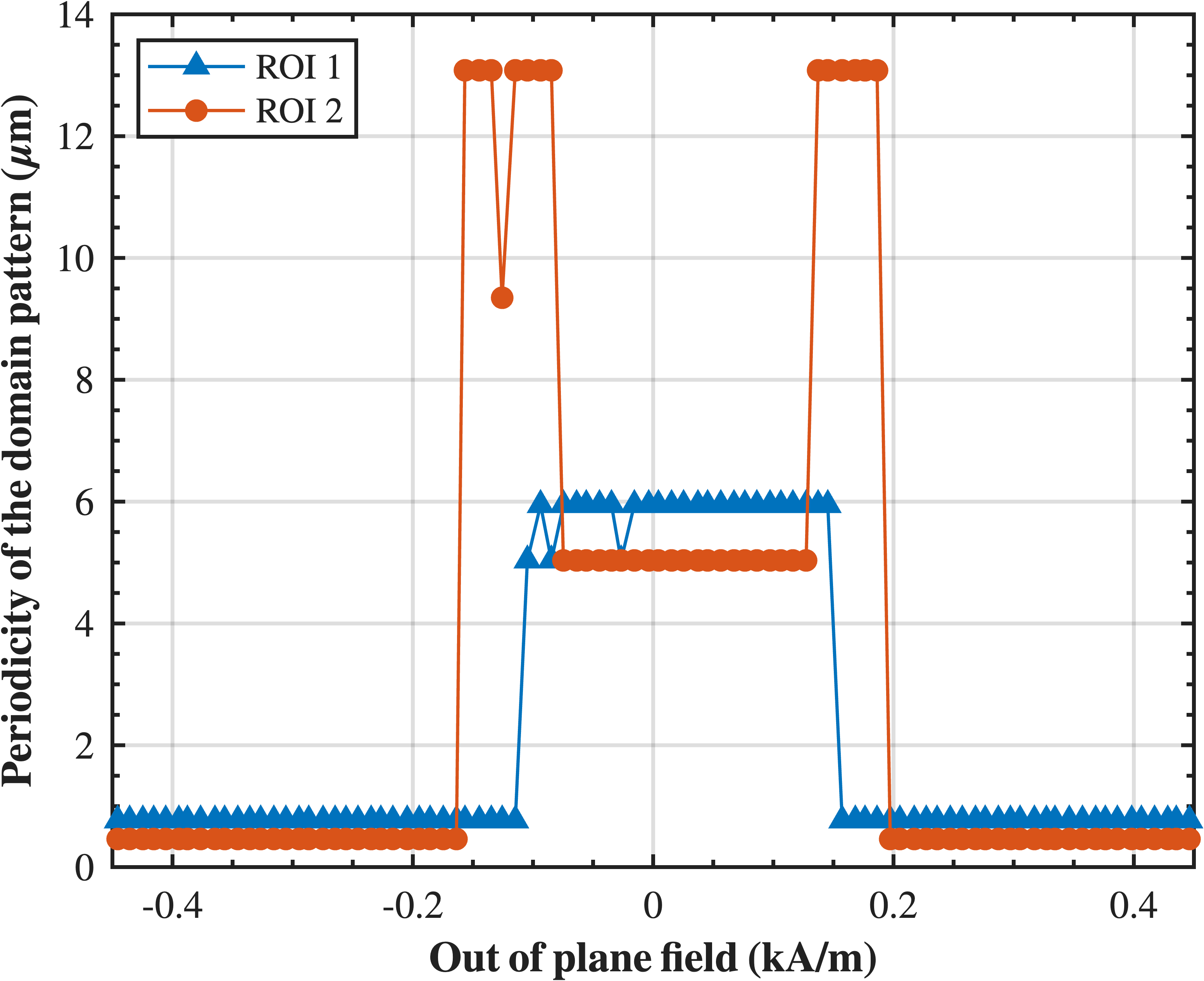}};
        \node[anchor=north west, xshift = -3mm, yshift= 2mm, fill=none, fill opacity=0.8, text opacity=1, font=\footnotesize, inner sep=2pt] at (img.north west) {(a)};
    \end{tikzpicture}
    
    \vspace{3mm} 
    
    % --- BOTTOM ROW: All MOKE images (b, c, d, e, f) side-by-side ---
    \begin{minipage}[t]{0.18\linewidth}\vspace{0pt}
        \centering
        \begin{tikzpicture}
            \node[inner sep=0pt, draw=black, thick] (img) {\includegraphics[width=\linewidth]{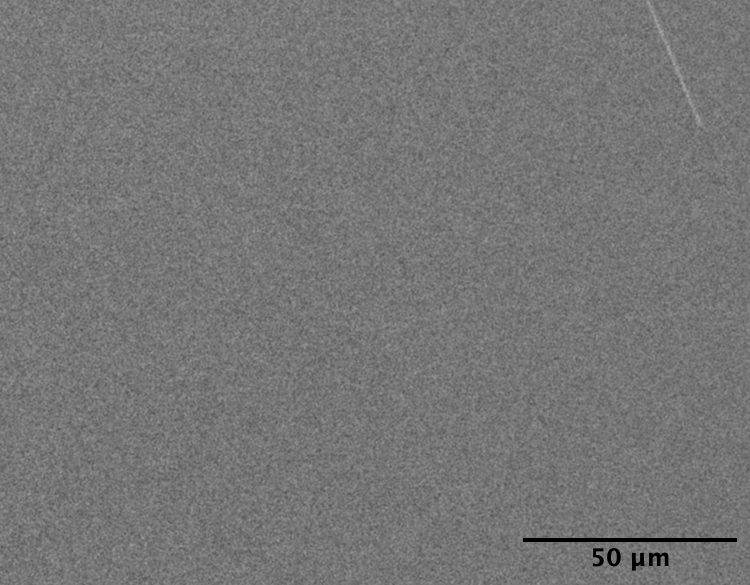}};
            \node[anchor=north west, fill=white, fill opacity=0.8, text opacity=1, font=\tiny, inner sep=1pt] at (img.north west) {(b)};
        \end{tikzpicture}
    \end{minipage}\hfill
    \begin{minipage}[t]{0.18\linewidth}\vspace{0pt}
        \centering
        \begin{tikzpicture}
            \node[inner sep=0pt, draw=black, thick] (img) {\includegraphics[width=\linewidth]{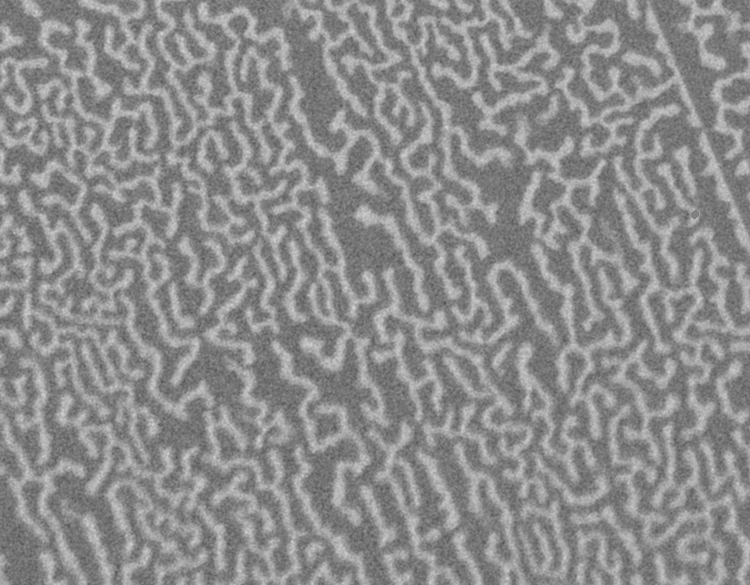}};
            \node[anchor=north west, fill=white, fill opacity=0.8, text opacity=1, font=\tiny, inner sep=1pt] at (img.north west) {(c)};
        \end{tikzpicture}
    \end{minipage}\hfill
    \begin{minipage}[t]{0.18\linewidth}\vspace{0pt}
        \centering
        \begin{tikzpicture}
            \node[inner sep=0pt, draw=black, thick] (img) {\includegraphics[width=\linewidth]{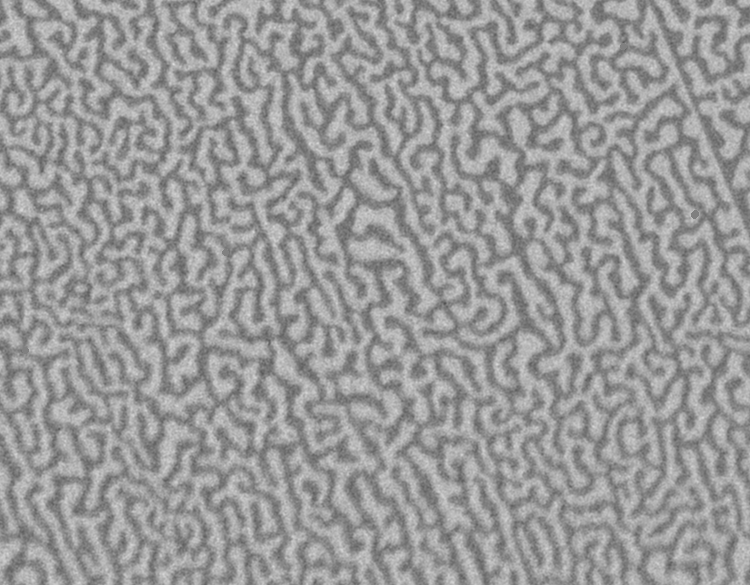}};
            \node[anchor=north west, fill=white, fill opacity=0.8, text opacity=1, font=\tiny, inner sep=1pt] at (img.north west) {(d)};
        \end{tikzpicture}
    \end{minipage}\hfill
    \begin{minipage}[t]{0.18\linewidth}\vspace{0pt}
        \centering
        \begin{tikzpicture}
            \node[inner sep=0pt, draw=black, thick] (img) {\includegraphics[width=\linewidth]{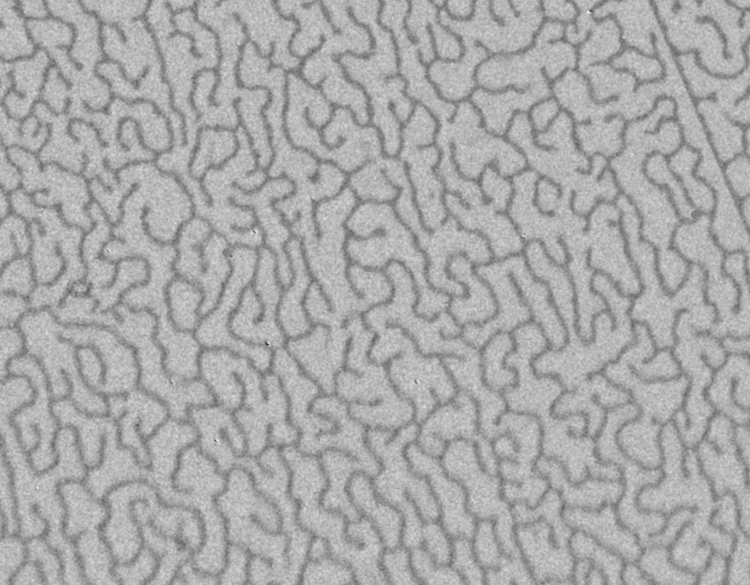}};
            \node[anchor=north west, fill=white, fill opacity=0.8, text opacity=1, font=\tiny, inner sep=1pt] at (img.north west) {(e)};
        \end{tikzpicture}
    \end{minipage}\hfill
    \begin{minipage}[t]{0.18\linewidth}\vspace{0pt}
        \centering
        \begin{tikzpicture}
            \node[inner sep=0pt, draw=black, thick] (img) {\includegraphics[width=\linewidth]{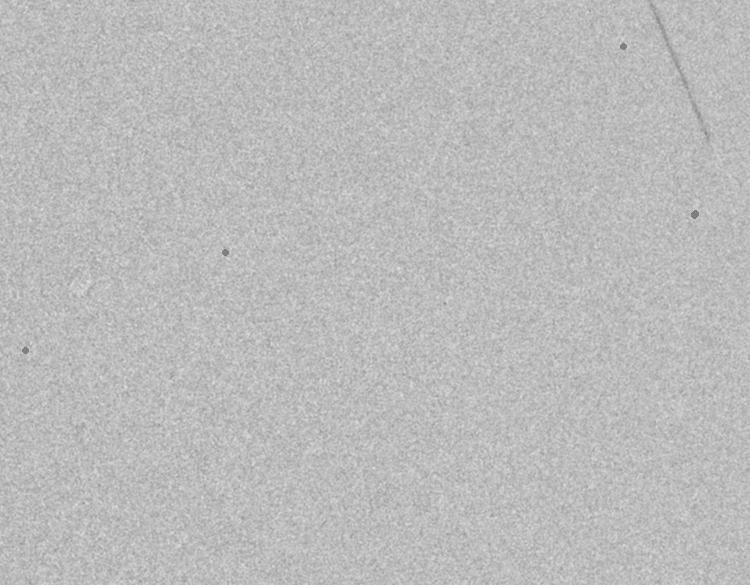}};
            
            % ROI 1 and ROI 2 boxes
            \node[draw=orange, thick, minimum width=15mm, minimum height=12mm, anchor=south, xshift=0mm, yshift=0.5mm] (roi1_box) at (img.south) {};
            % Text shifted slightly upward inside the box
            \node[text=orange, font=\bfseries\tiny, anchor=south, yshift=1mm, inner sep=0.2pt] at (roi1_box.south) {ROI 2};
            
            \node[draw=cyan, thick, minimum width=13mm, minimum height=8mm, anchor=north east, xshift=-0.2mm, yshift=-0.2mm] (roi2_box) at (img.north east) {};
            \node[text=cyan, font=\bfseries\tiny, anchor=center] at (roi2_box.center) {ROI 1};

            \node[anchor=north west, fill=white, fill opacity=0.8, text opacity=1, font=\tiny, inner sep=1pt] at (img.north west) {(f)};
        \end{tikzpicture}
    \end{minipage}
    
    \vspace{2mm}
    
    \caption{Magnetic domain-pattern periodicity as a function of the applied OOP field. 
    (a)~Evolution of the domain periodicity during an OOP field sweep from negative to positive values. The extracted data for ROIs 1 and 2 correspond to the distinct sample areas indicated in (f). Near zero external field, the periodicity of the domain pattern is $5.64\,\mu\mathrm{m}$. 
    (b)--(f)~MOKE images acquired at applied OOP fields of $-0.43$, $-0.06$, $0.02$, $0.13$, and $0.41\,\mathrm{kA/m}$, respectively. The scale bar in (b) applies to all images.}
    \label{fig4}
\end{figure}

The evolution of the domain-pattern periodicity as a function of the applied OOP field is plotted in Fig.~\ref{fig4}. A notable observation is the nearly constant periodicity of the domains over the range of $+H_c$ to $-H_c$ while the respective widths of the majority and minority domains vary. . Using $A_{\mathrm{ex}} = 27.5\,\mathrm{pJ/m}$~\cite{31Bhat2023}, $M_{\mathrm{ref}} = 1650\,\mathrm{kA/m}$~\cite{25Lavr2009}, and $K_{\mathrm{eff}} = 106.28\,\mathrm{kJ/m^3}$ [see Fig.~\ref{fig1}(d)], the measured zero-field periodicity of $5.6 \pm 0.2\,\mu\mathrm{m}$ yields a DMI magnitude of $0.95 \pm 0.16\,\mathrm{mJ/m^2}$. This substantially smaller DMI value as compared to edge nucleation and BLS (see DMI measurement section C)results from the conventional approximation of infinitely sharp DWs.

The sharp-wall approximation completely neglects the finite width over which spins progressively rotate across the DW. Consequently, it fails to account for the magnetostatic interactions of the resulting volume charges and their associated energy penalties. This omission significantly underestimates the DMI, yielding a value approximately $1.6$ times smaller than that obtained from BLS. To overcome this limitation, we employ an analytical model for the stray-field energy of parallel stripe domains in multilayers with perpendicular magnetic anisotropy that explicitly incorporates a finite DW width and a variable DW angle~\cite{33Leme2017,34Agra2019}. Within this refined framework, the DMI is evaluated by minimizing the total energy density:
\begin{equation}
\begin{split}
E_{\mathrm{tot}} &= \frac{2}{\lambda} \left[ \frac{2A_{\mathrm{ex}}}{\Delta} + 2K_u \Delta + \pi D \sin(\psi) \right] \\
&\quad + E_s(M_{\mathrm{s}}, t, \lambda, W, \Delta) + E_v(M_{\mathrm{s}}, t, \lambda, W, \Delta, \psi) \\
&\quad - M_{\mathrm{s}} \left( 1 - \frac{2W}{\lambda} \right) H_z ,
\end{split}
\label{equation2}
\end{equation}
\noindent where $\lambda$ is the domain periodicity, $W$ is the minority-domain width, $t$ is the film thickness, $\Delta$ is the DW width, $\psi$ is the variable DW angle, and $K_{\mathrm{eff}} = K_u - \frac{1}{2} \mu_0 M_{\mathrm{s}}^2$. Here, $E_s$ and $E_v$ represent the surface and volume stray-field energy densities, respectively (further details regarding the system of equations and series convergence are provided in Fig.~\ref{fig:S7} of the supporting information). The DMI magnitude is extracted by numerically minimizing $E_{\mathrm{tot}}$. Because our sample lies in the ultrathin-film limit and its DMI exceeds the threshold value [Eq.~(\ref{eq:S7})], we can assume pure N\'{e}el DWs with a width that is independent of the DMI [Eq.~(\ref{eq:S5})]. Under these assumptions and using the bulk exchange stiffness of $A_{\mathrm{ex}} = 27.5\,\mathrm{pJ/m}$~\cite{35Peri}, we obtain a DMI magnitude of $1.60 \pm 0.22\,\mathrm{mJ/m^2}$.

\subsection{Brillouin Light Scattering}
BLS probes the dynamical properties of magnetic materials by analyzing frequency shifts in inelastically scattered light. This technique provides direct insight into spin-wave excitations~\cite{4Kuep2023,23Kuep2024} and is widely used to characterize magnetic parameters and interfacial effects in ultrathin films. Specifically, BLS enables a direct measurement of the interfacial DMI by quantifying the frequency nonreciprocity of counterpropagating spin waves~\cite{23Kuep2024,36Belm2016}. In the Damon--Eshbach geometry, the presence of interfacial DMI induces an asymmetry between the Stokes and anti-Stokes spin-wave frequencies, providing a direct measure of the DMI strength~\cite{4Kuep2023,23Kuep2024,35Peri,36Belm2016,37Tacc2017}.

\begin{figure}[htbp]
    \centering
    \begin{minipage}[t]{0.48\linewidth}
        \centering
        \begin{tikzpicture}
            \node[inner sep=0pt] (img) {\includegraphics[width=\linewidth]{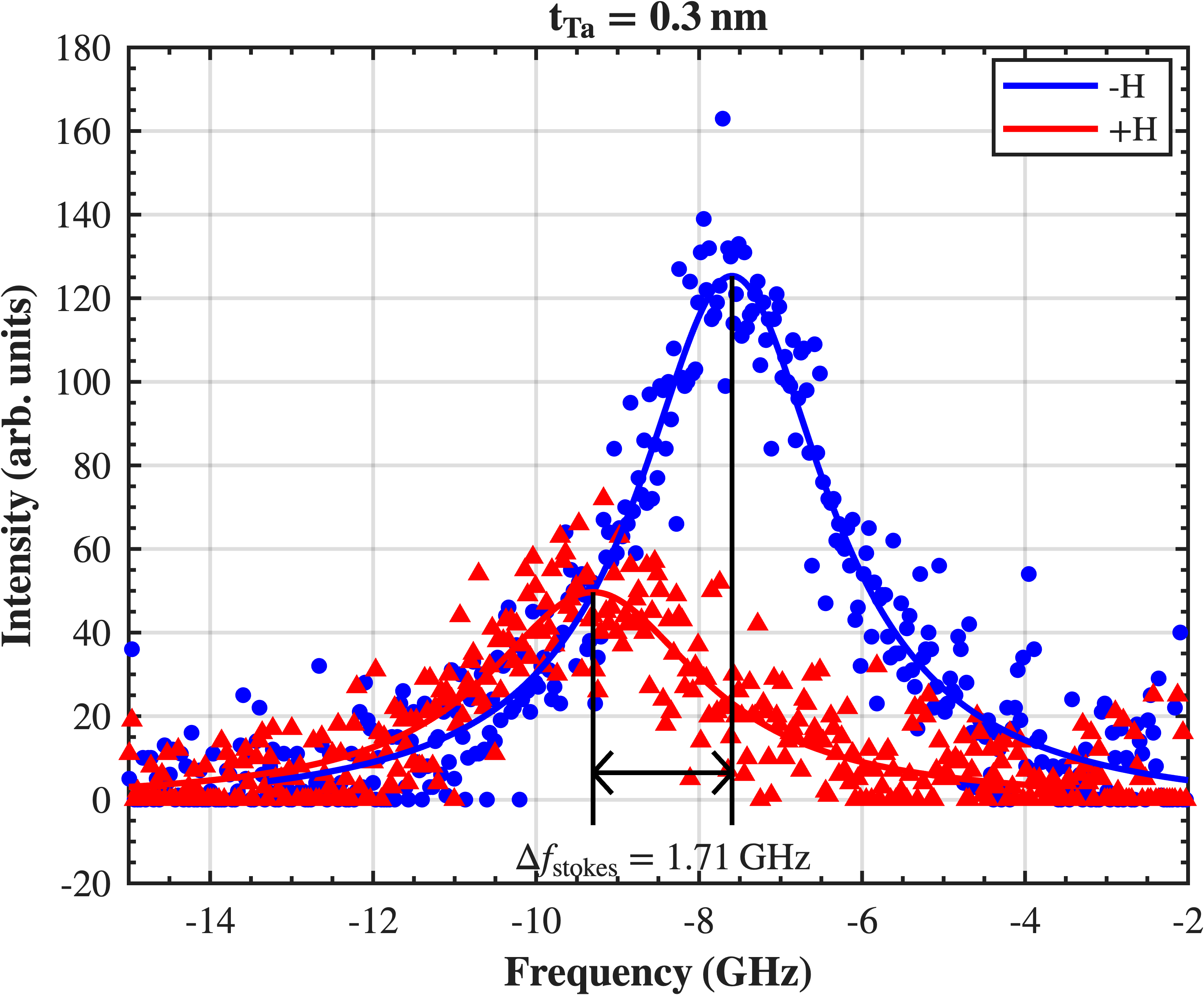}};
            \node[anchor=north west, xshift=-3mm, yshift=2mm, fill=none, fill opacity=0.5, text opacity=1, font=\scriptsize, inner sep=1.5pt] at (img.north west) {(a)};
        \end{tikzpicture}
    \end{minipage}\hfill
    \begin{minipage}[t]{0.48\linewidth}
        \centering
        \begin{tikzpicture}
            \node[inner sep=0pt] (img) {\includegraphics[width=\linewidth]{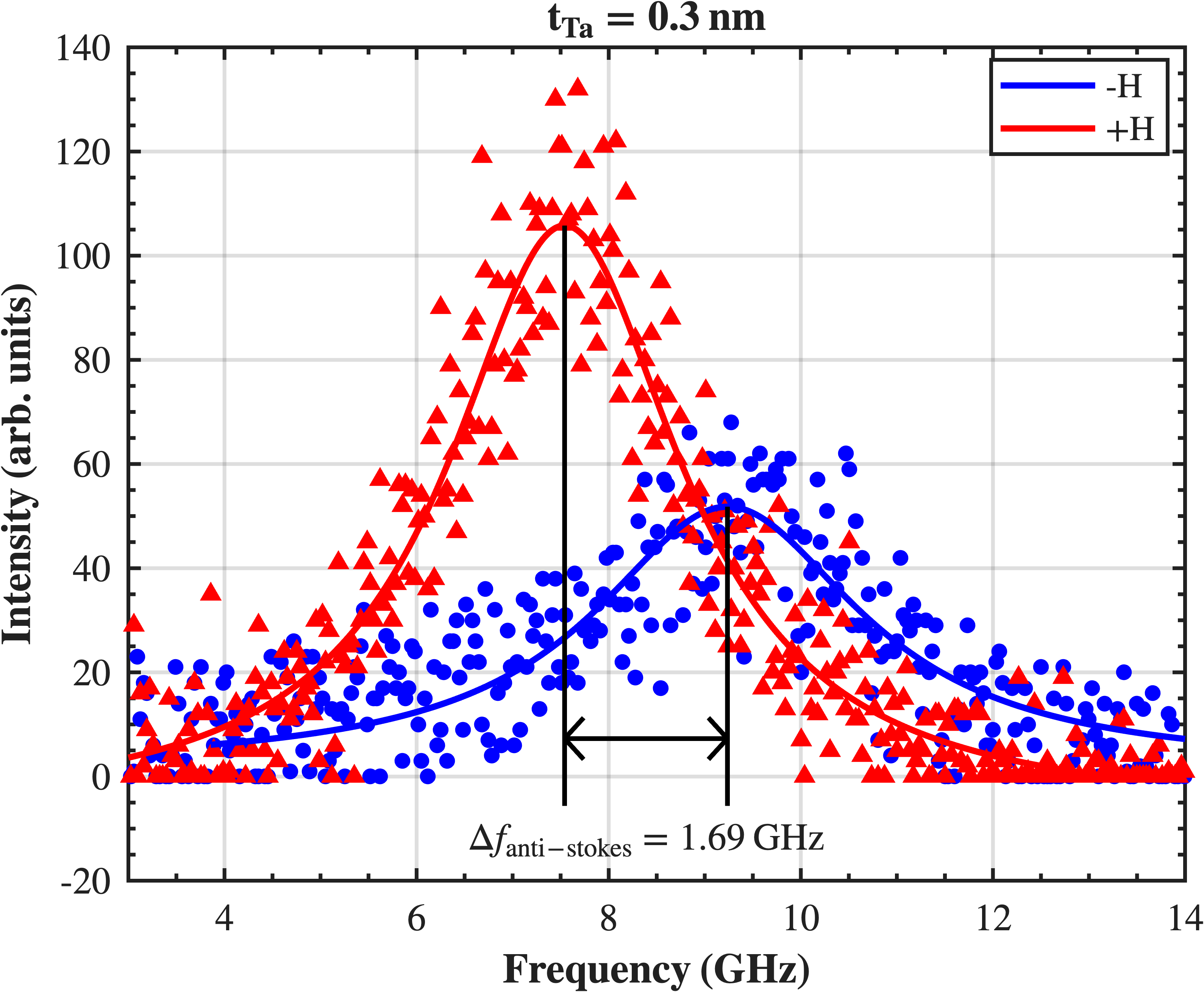}};
            \node[anchor=north west, xshift=-3mm, yshift=2mm, fill=none, fill opacity=0.5, text opacity=1, font=\scriptsize, inner sep=1.5pt] at (img.north west) {(b)};
        \end{tikzpicture}
    \end{minipage}
    
    \vspace{2mm}
    
    \caption{BLS spectra for the $0.3\,\mathrm{nm}$ Ta sample grown on a Si substrate, showing the (a)~Stokes and (b)~anti-Stokes peaks. Solid lines represent Lorentzian fits to the experimental data. Red triangles and blue circles denote measurements acquired under positive ($+H$) and negative ($-H$) IP saturation fields, respectively. The nonreciprocal frequency shift is defined as $\Delta f_{\mathrm{S/AS}} = \left|f_{\mathrm{S/AS}}(+H) - f_{\mathrm{S/AS}}(-H)\right|$, where $f_{\mathrm{S/AS}}(\pm H)$ corresponds to the Stokes (S) or anti-Stokes (AS) peak frequency at $\pm H$.}
    \label{fig5}
\end{figure}

To reduce phonon-related background signals and improve spectral resolution, BLS measurements were performed on identical magnetic multilayer stacks grown on silicon (Si) substrates rather than LiNbO$_{3}$. The Stokes and anti-Stokes peak frequency shifts were measured at room temperature under oppositely oriented IP saturation fields ($H = \pm 405\,\mathrm{kA/m}$). Figure~\ref{fig5} displays the fitted BLS spectra for the $0.3\,\mathrm{nm}$ Ta sample, while the spectra for the remaining samples are presented in Fig.~\ref{fig:bls} of the supporting information. This magnon frequency nonreciprocity is a direct signature of the DMI. The DMI magnitude is quantified from the frequency shift using $\Delta f = \frac{2|\gamma|D}{\pi M_{\mathrm{s}}}k_M$, where $k_M = \frac{2\pi}{\lambda}\sin\theta$ is the magnon wave vector, $\lambda = 532\,\mathrm{nm}$ is the laser wavelength, $\gamma = 176\,\mathrm{GHz/T}$ is the gyromagnetic ratio, and $\theta = 45^{\circ}$ is the incident angle. From the measured Stokes and anti-Stokes frequency shifts, we extract DMI magnitudes of $1.50 \pm 0.10\,\mathrm{mJ/m^2}$ and $1.49 \pm 0.10\,\mathrm{mJ/m^2}$, respectively. Notably, in this dispersion relation, the extracted DMI values depend \textit{only} on the saturation magnetization of the film, which is independently measured for our samples.

\section{Results}
\begin{figure}[htbp]
    \centering
    \begin{minipage}[t]{0.48\linewidth}
        \centering
        \begin{tikzpicture}
            \node[inner sep=0pt] (img) {\includegraphics[width=\linewidth]{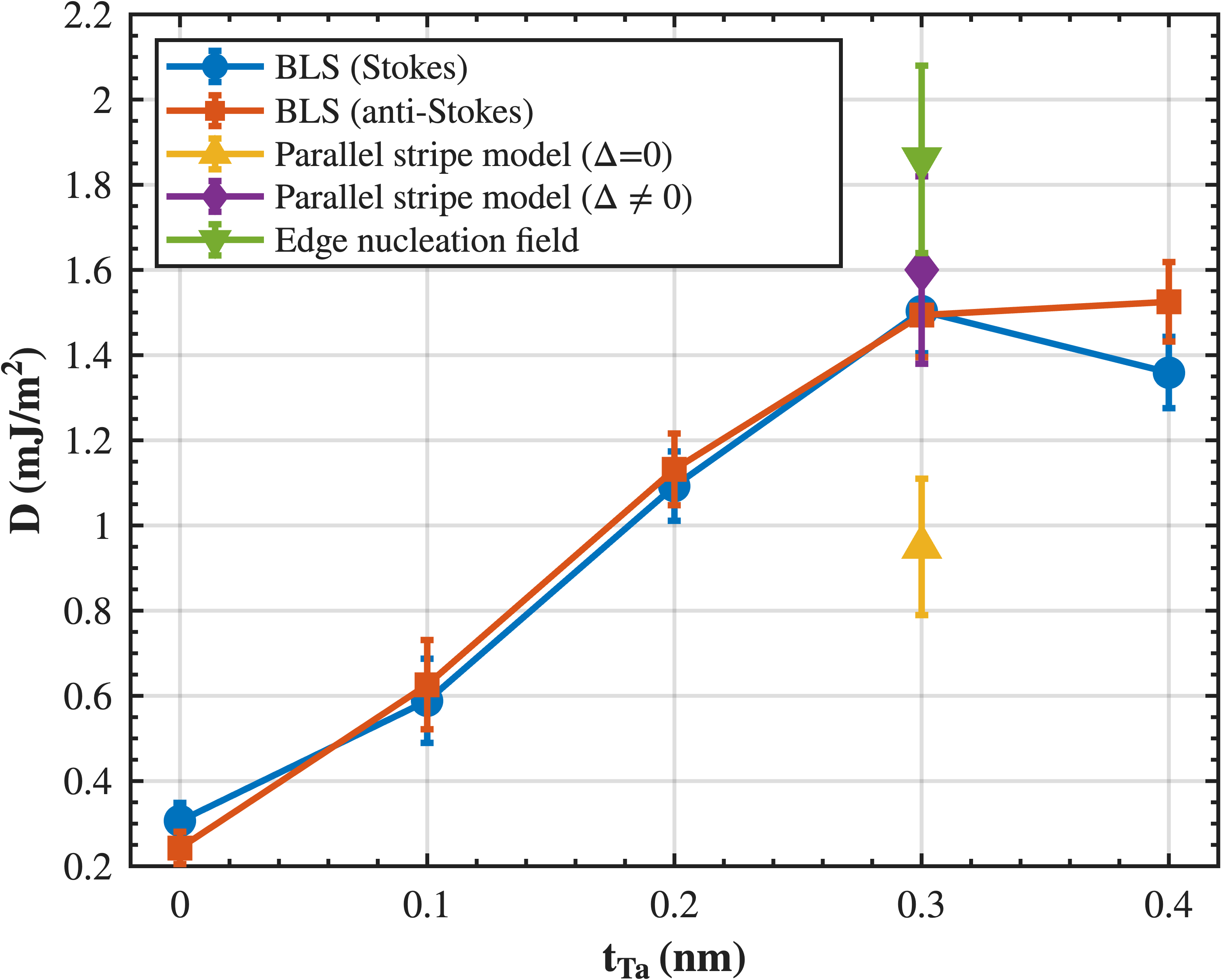}};
            \node[anchor=north west, xshift=-2mm, yshift=2mm, fill=none, fill opacity=0.5, text opacity=1, font=\scriptsize, inner sep=1.5pt] at (img.north west) {(a)};
        \end{tikzpicture}
    \end{minipage}\hfill
    \begin{minipage}[t]{0.48\linewidth}
        \centering
        \begin{tikzpicture}
            \node[inner sep=0pt] (img) {\includegraphics[width=\linewidth]{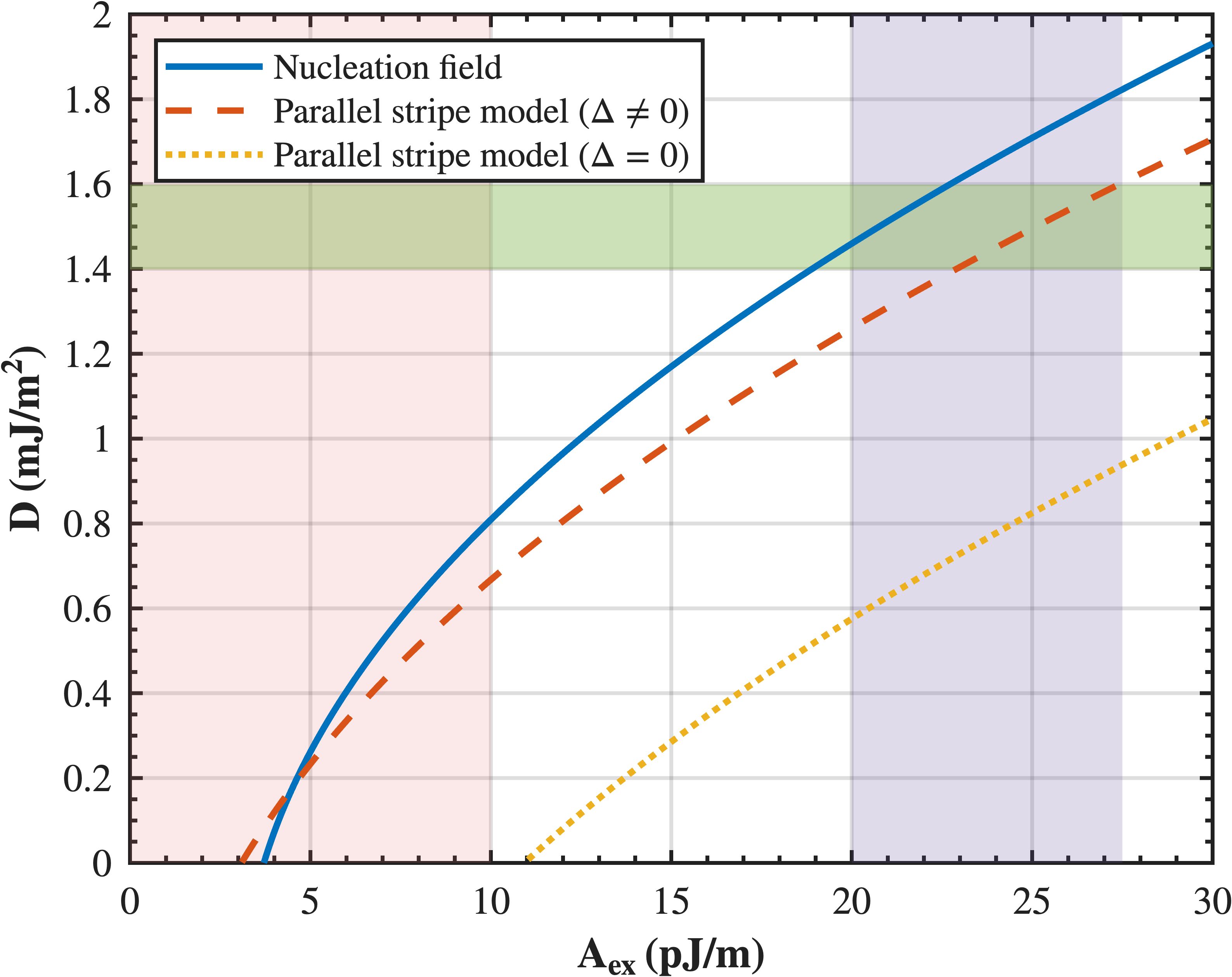}};
            \node[anchor=north west, xshift=-2mm, yshift=2mm, fill=none, fill opacity=0.5, text opacity=1, font=\scriptsize, inner sep=1.5pt] at (img.north west) {(b)};
        \end{tikzpicture}
    \end{minipage}
    
    \vspace{2mm}
    
    \caption{(a)~DMI values for the CoFeB heterostructures as a function of the Ta overlayer thickness ($t_{\mathrm{Ta}}$). Here, $t_{\mathrm{Ta}}=0\,\mathrm{nm}$ corresponds to the sample without Ta insertion. 
    (b)~Dependence of the extracted DMI value on the exchange stiffness ($A_{\mathrm{ex}}$) of CoFeB across the different measurement methods. The solid blue line represents the edge-nucleation field model, the dashed red line represents the parallel-stripe model assuming a finite DW width, and the dotted yellow line represents the parallel-stripe model assuming a zero DW width. The horizontal green line indicates the DMI value ($1.50 \pm 0.10\,\mathrm{mJ/m^2}$) obtained from BLS. The purple and orange shaded regions denote the reported exchange stiffness ranges for CoFeB in the ultrathin ($0.1$--$10\,\mathrm{pJ/m}$) and bulk ($20$--$27.5\,\mathrm{pJ/m}$) regimes, respectively.}
    \label{fig6}
\end{figure}

Figure~\ref{fig6}(a) shows the DMI values for the Pt/CoFeB/Ta samples obtained using the three independent methods. The DMI values extracted from BLS are plotted as a function of Ta thickness, where $t_{\mathrm{Ta}}=0\,\mathrm{nm}$ corresponds to the reference sample without submonolayer Ta insertion. The small net DMI observed in this reference sample, despite its nominally symmetric Pt interfaces, likely arises from differences in growth morphology and interfacial mixing~\cite{28Adhi2025}. The DMI increases monotonically with Ta thickness, from $0.27\,\mathrm{mJ/m^2}$ at $t_{\mathrm{Ta}}=0\,\mathrm{nm}$ to a maximum of approximately $1.50\,\mathrm{mJ/m^2}$ at $t_{\mathrm{Ta}}=0.3\,\mathrm{nm}$. Further increasing the Ta thickness is unlikely to enhance the DMI, as the trend approaches saturation above $0.3\,\mathrm{nm}$, a thickness comparable to the Ta unit-cell dimension of $0.33\,\mathrm{nm}$. For the $t_{\mathrm{Ta}}=0.3\,\mathrm{nm}$ sample, Fig.~\ref{fig6}(a) also compares the DMI values obtained from the edge-nucleation field analysis and both parallel-stripe models (assuming either a zero DW width or a more realistic finite DW width).

Assuming the bulk value of $A_{\mathrm{ex}}$, the three methods yield a range of DMI values. However, previous experiments on magnetic thin films have demonstrated that the exchange stiffness can vary widely and depends strongly on the interfacial composition~\cite{38Vaz2008,39Yast2019}. Unlike the BLS-derived values, both the edge-nucleation and parallel-stripe models depend explicitly on $A_{\mathrm{ex}}$, a parameter that is difficult to measure directly. For instance, recent studies on Ta/\allowbreak CoFeB($0.5\,\mathrm{nm}$)/\allowbreak MgO/\allowbreak CoFeB($t\,\mathrm{nm}$)/\allowbreak MgO heterostructures have shown that $A_{\mathrm{ex}}$ for Co$_{60}$Fe$_{20}$B$_{20}$ ranges from $0.5$ to $10\,\mathrm{pJ/m}$ for film thicknesses between $1.1$ and $1.9\,\mathrm{nm}$~\cite{35Peri}. This uncertainty in $A_{\mathrm{ex}}$ strongly influences the extracted DMI values. Figure~\ref{fig6}(b) illustrates the dependence of the DMI on $A_{\mathrm{ex}}$ for both the edge-nucleation and parallel-stripe models. Across the range of possible $A_{\mathrm{ex}}$ values, assuming a zero DW width leads to a wide disparity in the extracted DMI values, whereas accounting for a finite DW width significantly reduces this discrepancy.

\section{Conclusion}
We have demonstrated precise control of the interfacial DMI through subnanometer Ta insertion in Cr($2\,\mathrm{nm}$)/\allowbreak Ta($1\,\mathrm{nm}$)/\allowbreak Pt($2\,\mathrm{nm}$)/\allowbreak CoFeB($0.7\,\mathrm{nm}$)/\allowbreak Ta($t_{\mathrm{Ta}}$)/\allowbreak Pt($3\,\mathrm{nm}$) heterostructures. Because this system is extremely sensitive to the Ta overlayer thickness, the DMI increases nearly sixfold from approximately $0.27$ to $1.50\pm0.10\,\mathrm{mJ/m^2}$, as revealed by BLS. Simultaneously, the perpendicular magnetic anisotropy is minimized, which substantially reshapes the magnetic domain configurations. This behavior contrasts sharply with Au- and Cu-capped Co thin films, where the magnetic anisotropy peaks near a single monolayer of coverage~\cite{40Enge1993}. Structural characterizations of those noble-metal overlayers show negligible intermixing or strain. In contrast, the Ta/CoFeB interface exhibits pronounced intermixing~\cite{24Tani2017}, which reduces the effective magnetic thickness~\cite{13Shen2019}. Consequently, subtle interfacial modifications strongly affect the magnetic anisotropy, which in turn alters the DW structure and domain pattern. This sensitivity underscores the necessity of simultaneous control over competing magnetic energy contributions (rather than the DMI alone) in the engineering of chiral magnetic textures.

Accurately quantifying the DMI requires rigorous analysis and cross-validation among multiple characterization techniques. Here, we compare three complementary methods: BLS measurements, edge-nucleation field analysis, and magnetic stripe-domain width analysis. Together, these techniques provide qualitative and quantitative insights into both the magnitude and sign of the DMI, confirming a left-handed chirality. Among these methods, BLS offers the most direct determination of the DMI magnitude because it requires only the saturation magnetization and does not explicitly depend on the exchange stiffness. However, BLS requires specialized instrumentation, and extracting the sign of the DMI demands careful analysis~\cite{4Kuep2023}. Conversely, imaging chirality-dependent magnetization reversal via edge nucleation visually verifies the presence and sign of the DMI immediately. Determining its magnitude, however, requires field-dependent measurements alongside independent knowledge of the exchange stiffness, magnetic anisotropy, and saturation magnetization. Similarly, stripe-domain analysis has experimental constraints. It requires samples with well-established maze or stripe configurations and minimal pinning, which prevents its reliable application to the $0.1\,\mathrm{nm}$, $0.2\,\mathrm{nm}$, and $0.4\,\mathrm{nm}$ Ta overlayer samples (see Figs.~\ref{fig:S4} and~\ref{fig:S5}).

Beyond experimental constraints, our comparison shows that extracting reliable DMI values from domain-based models depends critically on theoretical assumptions. Both prior studies~\cite{33Leme2017,34Agra2019} and the present work highlight the limitations of the conventional sharp-wall approximation, which assumes a zero DW width. Because this approximation neglects the energy contributions of the wall's internal structure, it substantially underestimates the DMI relative to BLS. Incorporating a more realistic physical representation markedly improves quantitative agreement across the techniques. Specifically, this refined model accounts for finite DW widths, variable wall angles, and internal magnetostatic volume charges~\cite{33Leme2017}. Even within this improved framework, the fits remain sensitive to sample thickness and multiple micromagnetic parameters. The exchange stiffness $A_{\mathrm{ex}}$ is the primary source of uncertainty because it is difficult to measure independently. Notably, adopting a bulk-like value for $A_{\mathrm{ex}}$ yields the best agreement in the DMI values across all three methods. However, this result contrasts with recent reports of a reduced exchange stiffness in ultrathin CoFeB films~\cite{35Peri}. Resolving this discrepancy requires an independent determination of $A_{\mathrm{ex}}$ and a careful assessment of model-dependent uncertainties. Furthermore, the BLS and magnetic-domain measurements were performed on equivalent heterostructures grown on different substrates. Therefore, potential substrate-induced effects must also be considered when evaluating quantitative consistency.

Overall, these findings link precision interfacial engineering, competing magnetic interactions, and reliable DMI characterization in ultrathin magnetic heterostructures. We have demonstrated wide tunability of the DMI and cross-validated it using three complementary measurement techniques. Together, these results provide a robust reference for stabilizing chiral magnetic textures and refining the micromagnetic parameters needed for spintronic device design. Future studies should integrate independently determined exchange stiffness, substrate-matched characterization, and current-driven DW dynamics. Such investigations will be vital for translating these materials-level insights into quantitative predictions of device performance.

\begin{acknowledgments}
This work was supported by the National Science Foundation (NSF) under EPSCoR RII Track-1 Grant No. OIA-2044049 (EQUATE) and Division of Materials Research Grant No. 2439947. This research was performed in part in the Nebraska Nanoscale Facility: National Nanotechnology Coordinated Infrastructure and the Nebraska Center for Materials and Nanoscience (and/or NERCF), which are supported by the National Science Foundation under Award No. ECCS: 2025298 and the Nebraska Research Initiative. 
\end{acknowledgments}

%\section{Author Contributions}
%\textbf{Rabindra R Oliya}: Conceptualization (equal); Data curation (lead); Formal analysis (lead); Investigation (lead); Methodology (equal); Writing-original draft (equal); Writing- reviewing and editing (equal). \textbf{Christopher Keck}: Data curation (supporting). \textbf{Lia Compton}: Data curation (supporting); Formal analysis (supporting); Writing – review \& editing (supporting). \textbf{Jing Guo}: Data curation (supporting). \textbf{Kristen S. Buchanan}: Supervision (supporting);  Writing – review \& editing (supporting). \textbf{Robert Streubel}: Data curation (supporting); Formal analysis (supporting); Funding acquisition (supporting); Supervision (supporting); Writing – review \& editing (supporting). \textbf{Shireen Adenwalla}: Conceptualization (equal); Formal analysis (equal); Funding acquisition (lead); Methodology (equal);Resources (lead); Supervision (lead); Writing – original draft (equal); Writing – review and editing (equal). 

%%%%%%%%%%%%%%%%%%%%%%%%%%%%%%%%%%%%%%%%%%%%%%%%%%%%%%%%%%%%%%%%%%%%%
%% Appendix / Supporting Information
%%%%%%%%%%%%%%%%%%%%%%%%%%%%%%%%%%%%%%%%%%%%%%%%%%%%%%%%%%%%%%%%%%%%%
\appendix
\setcounter{figure}{0}
\renewcommand{\thefigure}{S\arabic{figure}}
\setcounter{equation}{0}
\renewcommand{\theequation}{S\arabic{equation}}

\section{Supporting Information}
\subsection{Sample Preparation}
The thin film hetero-structure is deposited using magnetron sputtering on both Si and $128^{\circ}$ Y-Cut LiNbO$_{3}$. The sputtering rates and corresponding power settings for each element in the hetero-structure were as follows: chromium (Cr) at $0.21$~\AA/s with $40$~W (DC), tantalum (Ta) at $0.13$~\AA/s with $40$~W (RF), Co$_{60}$Fe$_{20}$B$_{20}$ at $0.096$~\AA/s with $75$~W (RF), and platinum (Pt) at $0.27$~\AA/s with $20$~W (DC) at a base pressure of about $3\times 10^{-6}$~torr.   

\begin{figure}[htbp]
    \centering
    \includegraphics[width=0.48\linewidth]{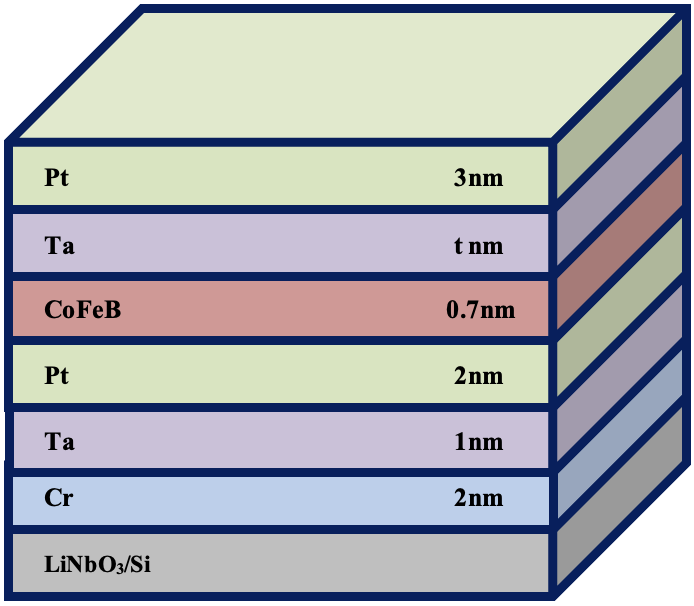}
    \caption{Schematic of the thin film heterostructure of Cr($2\,\mathrm{nm}$)/\allowbreak Ta($1\,\mathrm{nm}$)/\allowbreak Pt($2\,\mathrm{nm}$)/\allowbreak CoFeB($0.7\,\mathrm{nm}$)/\allowbreak Ta($t_{\mathrm{Ta}}$)/\allowbreak Pt($3\,\mathrm{nm}$). The numbers represent nominal film thicknesses, determined by the deposition rate.}
    \label{fig:S1}
\end{figure}

\subsection{Effective thickness of CoFeB}

\begin{figure}
    \centering
    \includegraphics[width=0.48\textwidth]{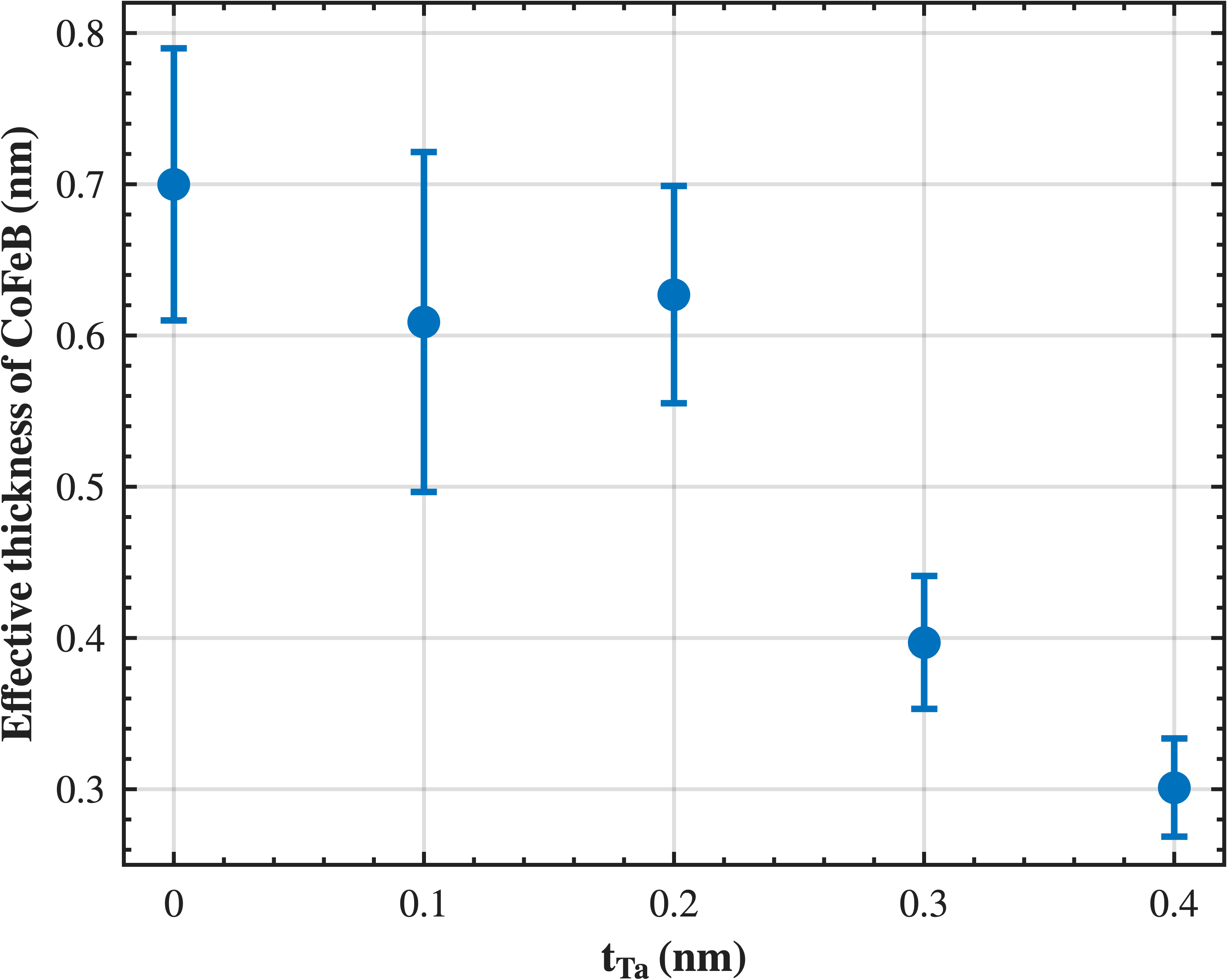}
    \caption{Assuming a constant value for the saturation magnetization, an estimate of the effective thickness of CoFeB as a function of the thickness of the Ta overlayer.}
    \label{fig:S2}
\end{figure}

Earlier measurements indicate that diffusion and alloying of Ta with CoFeB result in a magnetic dead layer at the Ta/CoFeB interface \cite{13Shen2019}. We estimated the effective thickness of the CoFeB to compensate for the dead layer contribution by $t_{\mathrm{eff}} = \frac{M_{\mathrm{measured}}}{M_{\mathrm{ref}}} \times t$, where $t$ is the thickness of CoFeB based on the sputtering calibration, $M_{\mathrm{measured}}$ is the measured saturation magnetization obtained from the measured magnetic moment $m = AM_{\mathrm{s}}(t - t_{\mathrm{d}})$, and $M_{\mathrm{ref}} = 1650\,\mathrm{kA/m}$ \cite{25Lavr2009} is the saturation magnetization for the Pt/CoFeB/Pt thin film, $A$ is the sample area, and $t_{\mathrm{d}}$ is the thickness of the magnetic dead layer. An estimate of the effective thickness of CoFeB as a function of the thickness of the Ta overlayer is shown in Fig~[\ref{fig:S2}].

\subsection{Kerr microscope image}

\begin{figure}[htbp]
    \centering
    \includegraphics[width=0.85\linewidth, keepaspectratio]{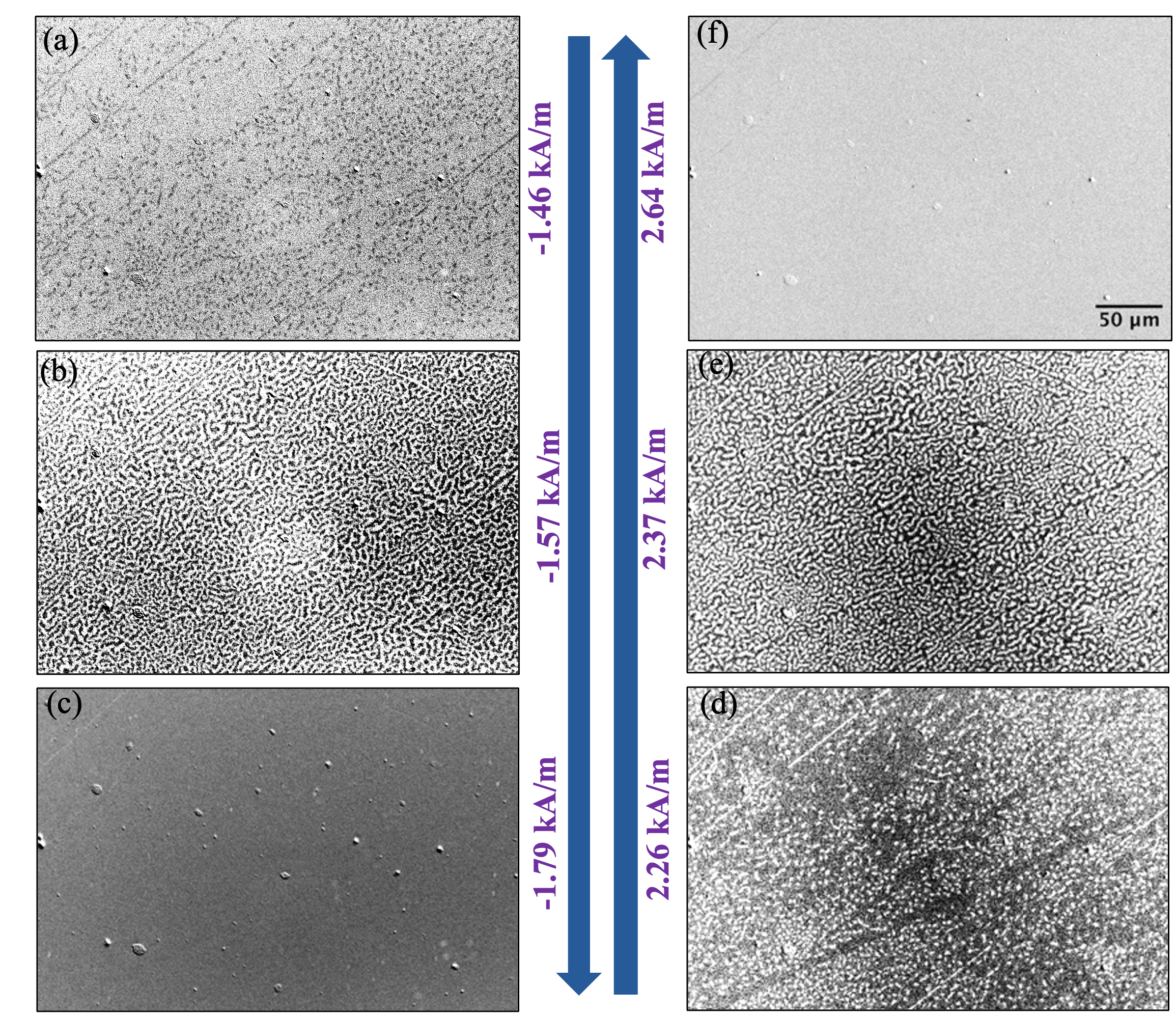}
    \caption{MOKE microscope images for the $t_{\mathrm{Ta}}=0.3\,\mathrm{nm}$ sample at different points along the hysteresis loop. The scale bar shown in (f) is the same for all the MOKE images.}
    \label{fig:S3}
\end{figure}

Fig~[\ref{fig:S3}] shows the MOKE microscope images of the $t_{\mathrm{Ta}}=0.3\,\mathrm{nm}$ sample at different points along the hysteresis loop. Bright areas represent regions with up magnetization, while the darker areas indicate down magnetization. Starting at positive saturation, images (a), (b), and (c) show the evolution of the domain structure as the negative OOP field is increased. At the corner of the hysteresis loop (see Fig~[\ref{fig1}(b)]), magnetic bubble domains with down magnetization nucleate [Fig~\ref{fig:S3}(a)]. With increasing OOP field, the magnetic bubble domains grow into a labyrinth domain [Fig~\ref{fig:S3}(b)] and then saturate [Fig~\ref{fig:S3}(c)]. Similar behavior is observed when we start with down magnetization and reverse the OOP field [Fig~\ref{fig:S3}(d)-\ref{fig:S3}(f)]. The $t_{\mathrm{Ta}}=0.1\,\mathrm{nm}$ and $t_{\mathrm{Ta}}=0.2\,\mathrm{nm}$ samples do not show magnetic bubble domains at nucleation. Magnetic domains are not observed in the $t_{\mathrm{Ta}}=0.4\,\mathrm{nm}$ sample within the resolution limit of our MOKE microscope. Taken together, these indicate the extreme sensitivity of domain formation to the Ta thickness.

\begin{figure}[htbp]
    \centering
    \includegraphics[width=0.85\linewidth, keepaspectratio]{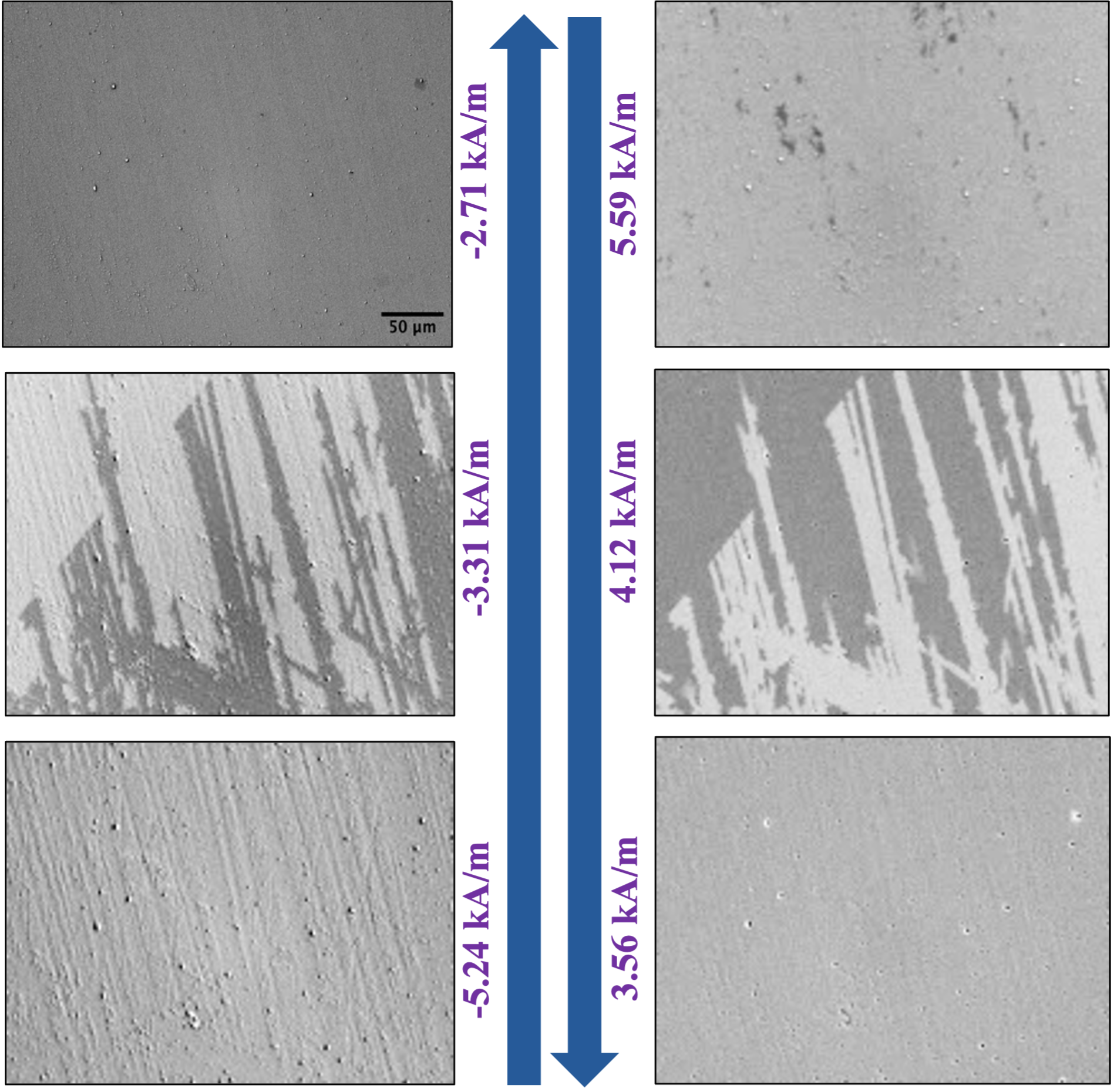}
    \caption{MOKE microscope image with $t_{\mathrm{Ta}} = 0.1\,\mathrm{nm}$ sample at different points along the hysteresis loop.}
    \label{fig:S4}
\end{figure}

\begin{figure}[htbp]
    \centering
    \includegraphics[width=0.85\linewidth, keepaspectratio]{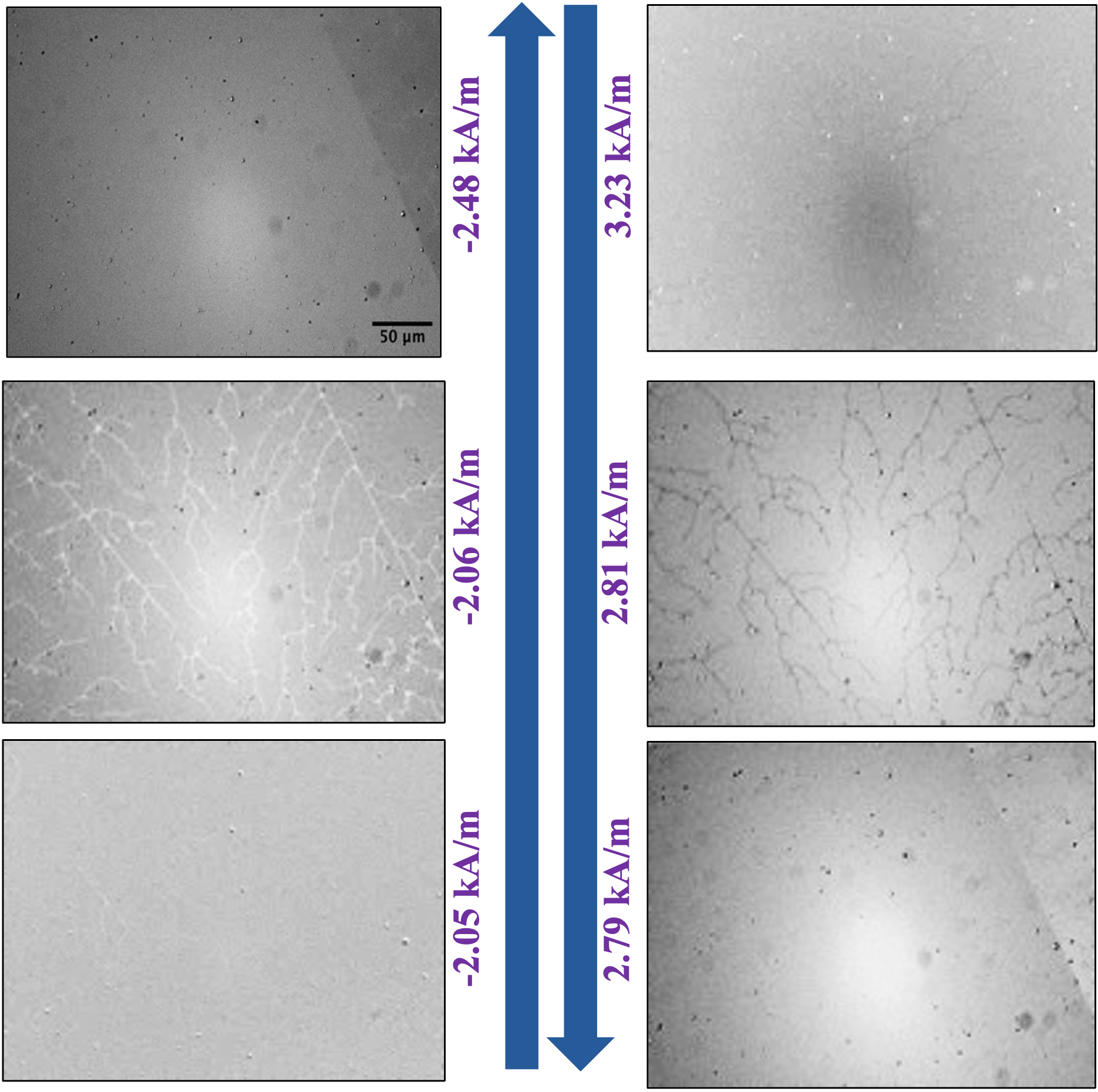}
    \caption{MOKE microscope image with $t_{\mathrm{Ta}} = 0.2\,\mathrm{nm}$ sample at different points along the hysteresis loop.}
    \label{fig:S5}
\end{figure}

\pagebreak
\subsection{Edge nucleation field}
The equation for the edge nucleation is given by \cite{32Sara2010}:

\begin{equation}
\begin{split}
  \mu_0 H_{n} &= \frac{2\pi A_{\mathrm{ex}} t}{p k_{\mathrm{B}} T} \left(1 - \frac{D}{D_{\mathrm{c}}}\right)^2 \mu_0 H_k \\
  &\quad \times \left(1 \mp \frac{\mu_0 H_x}{\mu_0 H_k \left(1 - \frac{D}{D_{\mathrm{c}}}\right)}\right)^2 \\
  &= \frac{\alpha}{H_k}H_x^2 \mp 2\alpha\beta H_x + \alpha \beta^2H_k
\end{split}
\label{eq:S1}
\end{equation}

where $\alpha = \frac{2\pi t_{\mathrm{eff}}}{p k_{\mathrm{B}} T}A_{\mathrm{ex}}$, $p = \ln\left(\frac{\tau}{\tau_0}\right) = \ln\left(\frac{100\,\mathrm{ms}}{1\,\mathrm{ns}}\right)$, $\beta = 1 - \frac{D}{D_{\mathrm{c}}}$, $D_{\mathrm{c}} =\frac{4}{\pi}\sqrt{A_{\mathrm{ex}}K_{\mathrm{eff}}}$, and $H_k = \frac{2K_{\mathrm{eff}}}{\mu_0 M_{\mathrm{s}}}$. By fitting a quadratic equation for $H_n$ as a function of the in-plane field $H_x$, we obtain the DMI.
 
\[ A = \frac{\alpha}{H_k} \]

\begin{equation}
B = \mp 2\alpha \beta \quad \Longrightarrow \quad D = D_{\mathrm{c}} \left[ 1 \mp \frac{B}{2\alpha} \right]
\label{eq:S2}
\end{equation}

\begin{equation}
C = \alpha\beta^2H_k \quad \Longrightarrow \quad D = D_{\mathrm{c}} \left[ 1 - \sqrt{\frac{C}{\alpha H_k} }\right]
\label{eq:S3}
\end{equation}
where $A$, $B$, and $C$ are fitting parameters.

\subsection{Parallel stripe model with finite domain wall width}

\begin{figure}[htbp]
    \centering
    \begin{minipage}[t]{0.48\linewidth}\vspace{0pt}
        \centering
        \begin{tikzpicture}
            \node[inner sep=0pt] (img) {\includegraphics[width=\linewidth]{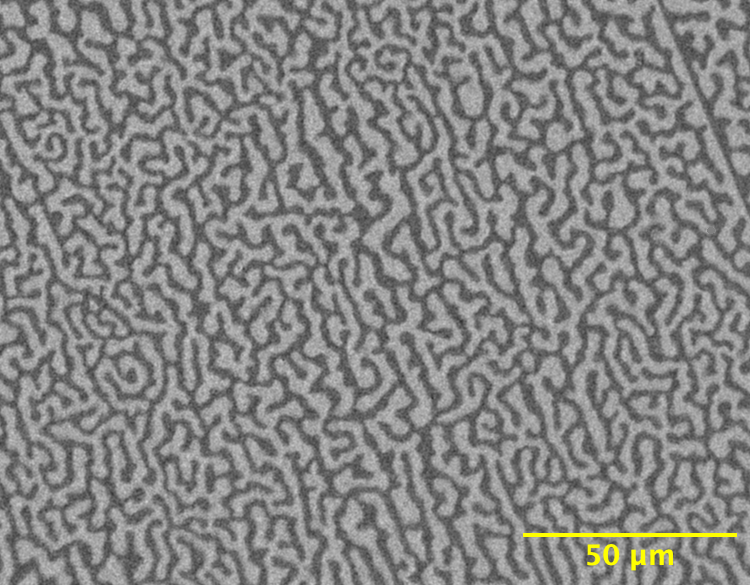}};
            \node[anchor=north west, xshift=-2mm, yshift = 3mm, fill=none, fill opacity=0.8, text opacity=1, font=\footnotesize, inner sep=1.5pt] at (img.north west) {(a)};
            
            % --- Added x and y axis at the bottom left ---
            \begin{scope}[shift={(img.south west)}, xshift=2mm, yshift=2mm]
                \draw[->, thick, white] (0,0) -- (0.6,0) node[right, font=\footnotesize] {$x$};
                \draw[->, thick, white] (0,0) -- (0,0.6) node[above, font=\footnotesize] {$y$};
            \end{scope}
            % ---------------------------------------------
            
        \end{tikzpicture}
    \end{minipage}\hfill
    \begin{minipage}[t]{0.48\linewidth}\vspace{0pt}
        \centering
        \begin{tikzpicture}
            \node[inner sep=0pt] (img) {\includegraphics[width=\linewidth]{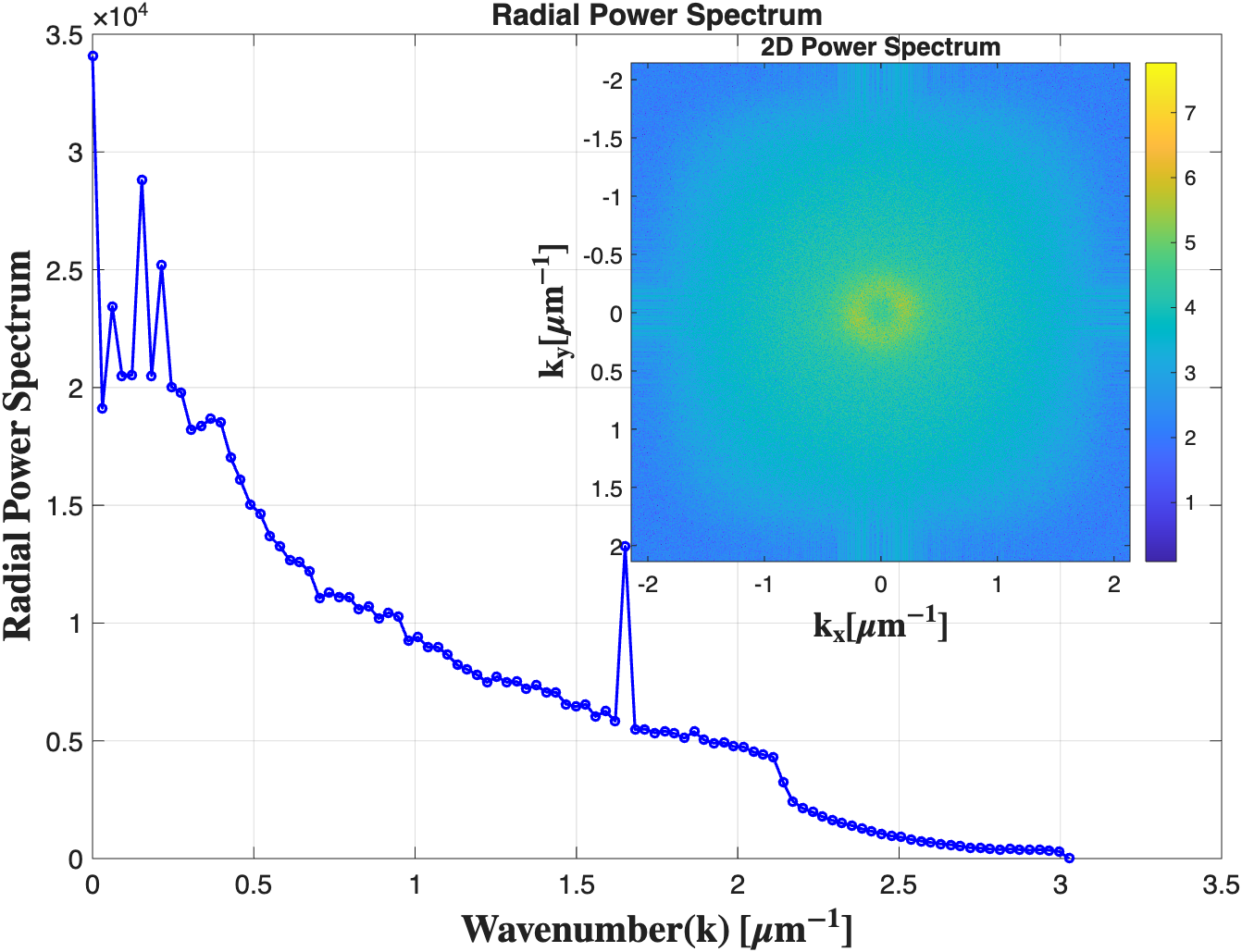}};
            \node[anchor=north west, xshift=-2mm, yshift = 3mm, fill=none, fill opacity=0.8, text opacity=1, font=\footnotesize, inner sep=1.5pt] at (img.north west) {(b)};
        \end{tikzpicture}
    \end{minipage}

    \vspace{4mm} 

    \begin{minipage}[t]{0.85\linewidth}\vspace{0pt}
        \centering
        \begin{tikzpicture}
            \node[inner sep=0pt] (img) {\includegraphics[width=\linewidth]{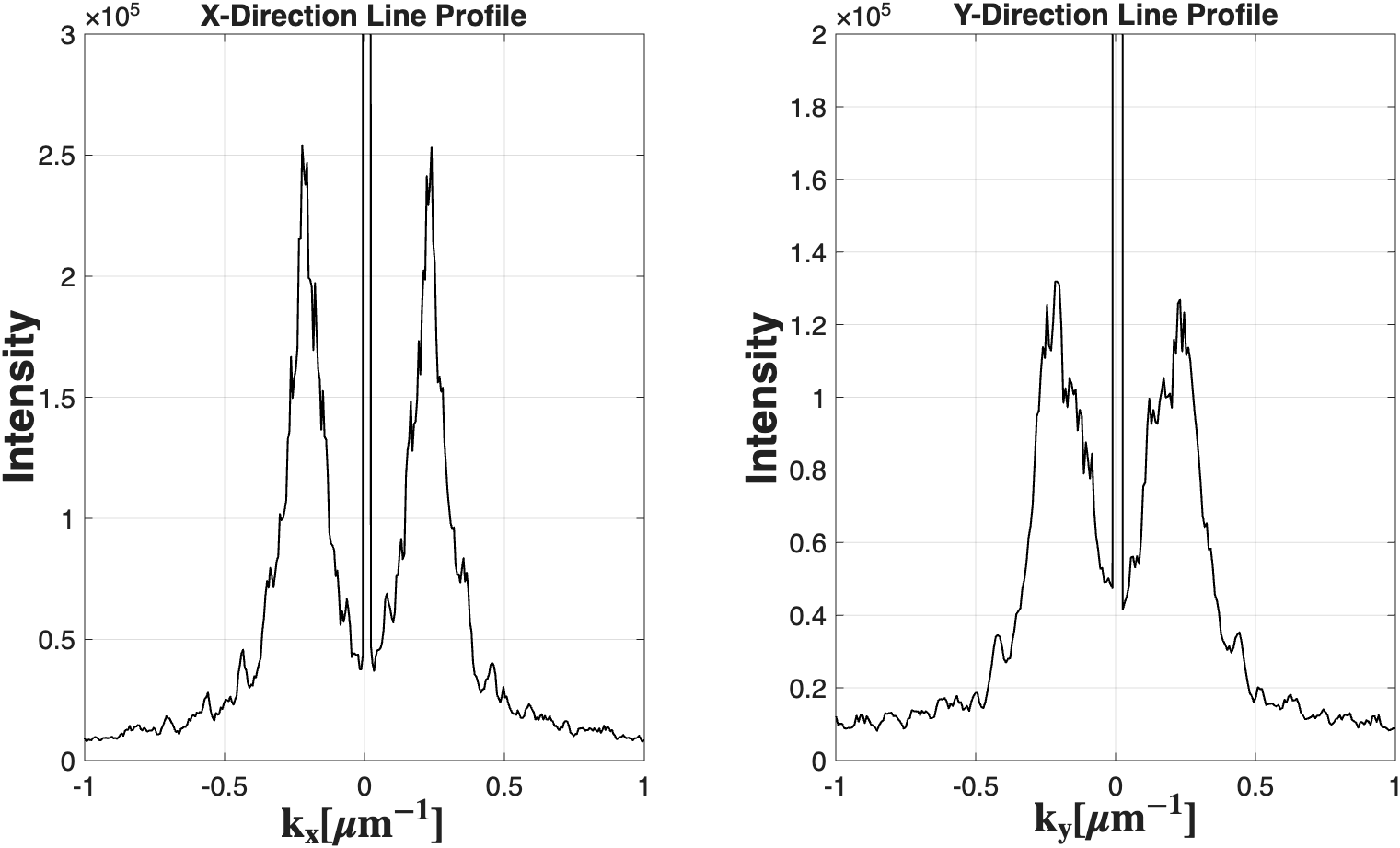}};
            \node[anchor=north west, xshift=-2mm, yshift = 3mm, fill=none, fill opacity=0.8, text opacity=1, font=\footnotesize, inner sep=1.5pt] at (img.north west) {(c)};
        \end{tikzpicture}
    \end{minipage}

    \caption{(a) MOKE image of $t_{\mathrm{Ta}}=0.3\,\mathrm{nm}$ sample at $0.02\,\mathrm{kA/m}$ OOP field. (b) The radial power spectrum of the 2D FFT of the image and the wavenumber $k=\sqrt{k_x^2 + k_y^2}$. The periodicity of the magnetic domain is $5.6 \pm 0.2\,\mu\mathrm{m}$. The inset displays the corresponding 2D power spectrum of the Fourier transformed image. (c) Line scans along $k_x$ and $k_y$ directions. The intensity is stronger along $k_x$, indicating the domain walls are preferentially oriented parallel to the y-axis.}
    \label{fig:S6}
\end{figure}

\noindent An accurate analytical model for the stray field energy of parallel stripe domains in multilayer films with perpendicular magnetic anisotropy, considering the effects of finite domain wall width and variable domain wall angle is given in \cite{33Leme2017}. Based on this model, the DMI of the thin film stripes is estimated by minimizing the total energy

\begin{equation}
\begin{split}
E_{\mathrm{tot}} &= \frac{2}{\lambda} \left[ \frac{2A_{\mathrm{ex}}}{\Delta} + 2K_u \Delta + \pi D \sin(\psi) \right] \\
&\quad + E_s(M_s, t, \lambda, W, \Delta) + E_v(M_s, t, \lambda, W, \Delta, \psi) \\
&\quad - M_s \left( 1 - \frac{2W}{\lambda} \right) H_z
\end{split}
\label{eq:S4}
\end{equation}

where $\lambda$ is the domain periodicity, $W$ is the minority domain width, $t$ is the film thickness, and $\Delta$ is the domain wall width given by

\begin{equation}
\lim_{t \to 0} \Delta = \sqrt{\frac{A_{\mathrm{ex}}}{K_{\mathrm{eff}}}} - \frac{\mu_0 M_{\mathrm{s}}^2 t}{4\pi K_{\mathrm{eff}}}
\label{eq:S5}
\end{equation}

\begin{equation}
\lim_{t \to \infty} \Delta = \sqrt{\frac{A_{\mathrm{ex}}}{K_u + \frac{1}{2}\mu_0 M_{\mathrm{s}}^2\sin^2(\psi)}},
\label{eq:S6}
\end{equation}

with effective anisotropy $K_{\mathrm{eff}} = K_u - \frac{1}{2}\mu_0 M_{\mathrm{s}}^2$, 

\begin{equation}
\begin{split}
\sin(\psi) &= 
\begin{cases} 
-\frac{D}{D_{\mathrm{thr}}} &;\, |D| < D_{\mathrm{thr}} \\ 
-\mathrm{sgn}(D) &;\, |D| \ge D_{\mathrm{thr}} 
\end{cases} \\
D_{\mathrm{thr}} &= \frac{2\mu_0 M_{\mathrm{s}}^2}{\frac{\pi^2}{t \ln(2)} + \pi \sqrt{\frac{K_u + \frac{1}{2}\mu_0 M_{\mathrm{s}}^2}{A_{\mathrm{ex}}}}}
\end{split}
\label{eq:S7}
\end{equation}

surface stray field energy 
\begin{equation*}
\begin{split}
E_s &= \frac{\mu_0 M_{\mathrm{s}}^2}{2}\left(1 - \frac{2W}{\lambda}\right)^2 \\
&\quad + \frac{2\pi \mu_0 M_{\mathrm{s}}^2 \Delta^2}{\lambda t} \sum_{n=1}^{\infty} \left[ \frac{\sin^2\left(\frac{n\pi W}{\lambda}\right)}{\sinh^2\left(\frac{\pi^2 n \Delta}{\lambda}\right)} \frac{1 - \exp\left(-\frac{2\pi n t}{\lambda}\right)}{n} \right]
\end{split}
\end{equation*}

and volume stray field energy 
\begin{equation*}
\begin{split}
E_v &= \frac{2\pi \mu_0 M_{\mathrm{s}}^2 \Delta^2 \sin^2(\psi)}{\lambda t} \\
&\quad \times \sum_{n=1}^{\infty} \left[ \frac{\sin^2\left(\frac{n\pi W}{\lambda}\right)}{\cosh^2\left(\frac{\pi^2 n \Delta}{\lambda}\right)} \frac{\exp\left(-\frac{2\pi n t}{\lambda}\right) + \frac{2\pi n t}{\lambda} - 1}{n} \right].
\end{split}
\end{equation*}

For $W = \lambda/2$ and $B_z = 0$, equation~[\ref{eq:S4}] reduces to

\begin{equation}
\begin{split}
E_{\mathrm{tot}} &= \frac{2}{\lambda} \left[ \frac{2A_{\mathrm{ex}}}{\Delta} + 2K_u \Delta - \pi D \right] \\
&\quad + E_s(M_s, t, \lambda, W, \Delta) + E_v(M_s, t, \lambda, W, \Delta)
\end{split}
\label{eq:S8}
\end{equation}

where

\begin{equation}
\begin{split}
E_s &= \frac{2\pi \mu_0 M_{\mathrm{s}}^2 \Delta^2}{\lambda t} \\
&\quad \times \sum_{n=1}^{\infty} \left[ \frac{\sin^2\left(\frac{n\pi}{2}\right)}{\sinh^2\left(\frac{\pi^2 n \Delta}{\lambda}\right)} \frac{1 - \exp\left(-\frac{2\pi n t}{\lambda}\right)}{n} \right],
\end{split}
\label{eq:S9}
\end{equation}

and

\begin{equation}
\begin{split}
E_v &= \frac{2\pi \mu_0 M_{\mathrm{s}}^2 \Delta^2 \sin^2(\psi)}{\lambda t} \\
&\quad \times \sum_{n=1}^{\infty} \left[ \frac{\sin^2\left(\frac{n\pi}{2}\right)}{\cosh^2\left(\frac{\pi^2 n \Delta}{\lambda}\right)} \frac{\exp\left(-\frac{2\pi n t}{\lambda}\right) + \frac{2\pi n t}{\lambda} - 1}{n} \right].
\end{split}
\label{eq:S10}
\end{equation}

\begin{figure}[htbp]
    \centering
    \includegraphics[width=0.85\linewidth]{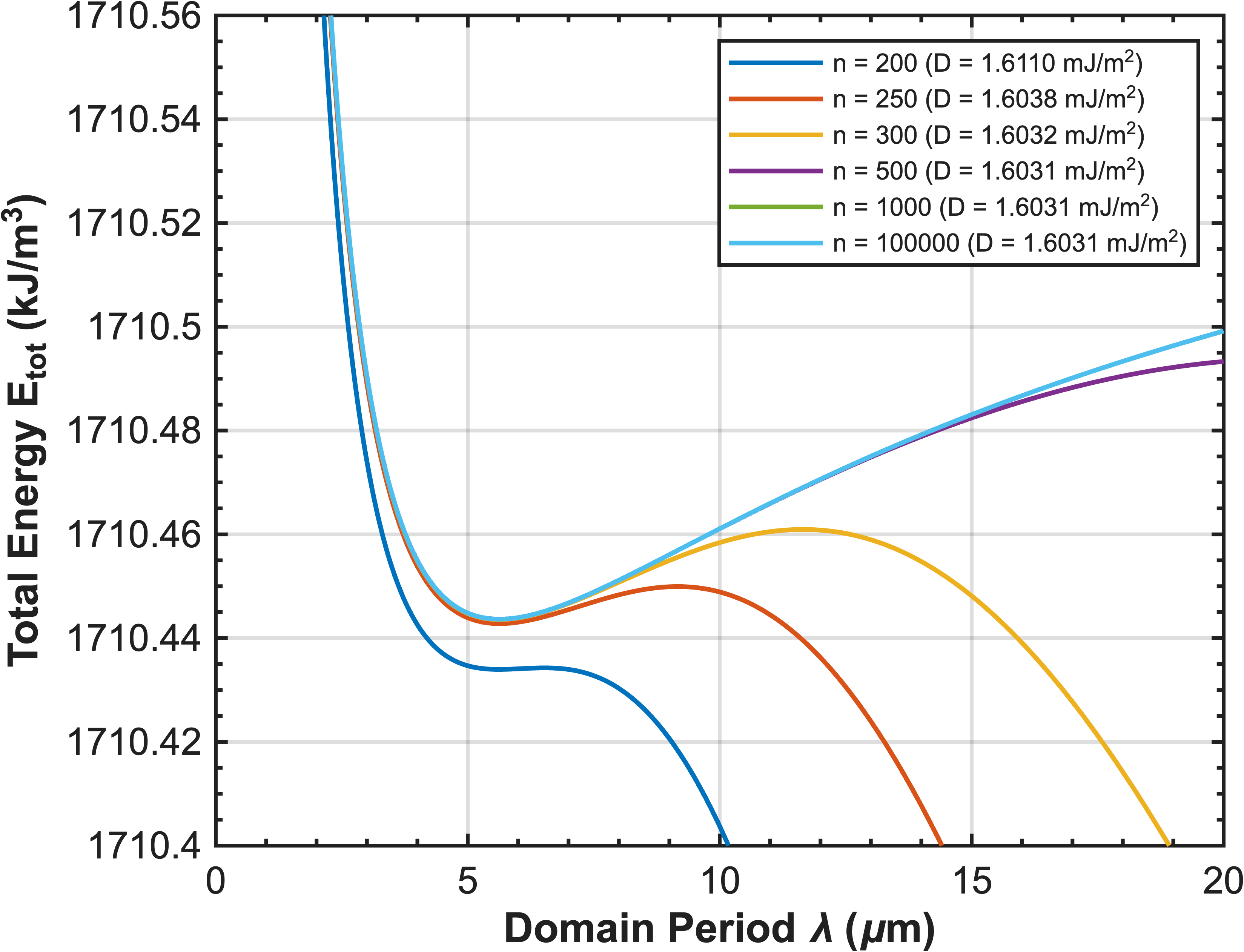}
    \caption{Total energy versus domain periodicity for various truncation limits ($n = 200, 250, 300, 500, 1000, 100000$) of the Fourier sum in equations~[\ref{eq:S9}] and [\ref{eq:S10}], ensuring the convergence of the series within the tolerance level ($10^{-15}$). The green ($n = 1000$) and sky blue ($n = 100000$) curves are superimposed. The energy is minimized at $\lambda_{\mathrm{exp}} = 5.64\,\mu\mathrm{m}$. The exchange parameter used is $A_{\mathrm{ex}} = 27.5\,\mathrm{pJ/m}$. The converged value of $D$ is $1.60\,\mathrm{mJ/m^2}$.}
    \label{fig:S7}
\end{figure}

\noindent For a $0.3\,\mathrm{nm}$ Ta dusting sample with $M_{\mathrm{s}} = 1650\,\mathrm{kA/m}$, $K_{\mathrm{eff}} = 106.28\,\mathrm{kJ/m^3}$, $A_{\mathrm{ex}} = 27.5\,\mathrm{pJ/m}$, and $t = 0.397\,\mathrm{nm}$. With these values $\Delta = 9.6\,\mathrm{nm}$, $\frac{\lambda}{\Delta} \approx 10^3$, and since $|D_{\mathrm{BLS}}| = 1.50\,\mathrm{mJ/m^2}$ and $D_{\mathrm{thr}} = 0.18\,\mathrm{mJ/m^2}$, $\sin(\psi) = -\mathrm{sgn}(D)$, the optimized DMI is obtained in such a way that the experimentally measured domain period $\lambda_{\mathrm{exp}}$ corresponds to a minimum of the total energy. In the $D \ge D_{\mathrm{thr}}$ regime, the domain wall becomes purely N\'{e}el and variations in DMI have no effect on the domain wall width ($\Delta$). Thus the energy is minimized with a single variable $\lambda$. The convergence of the equations [\ref{eq:S9}] and [\ref{eq:S10}] is shown in Fig [\ref{fig:S7}].

\pagebreak
\subsection{BLS measurement}
\begin{figure}[ht]
    \centering
    % --- ROW 1 (0.0 nm) ---
    \begin{minipage}[t]{0.48\linewidth}
        \centering
        \begin{tikzpicture}
            \node[inner sep=0pt] (img) {\includegraphics[width=\linewidth, height=0.12\textheight, keepaspectratio]{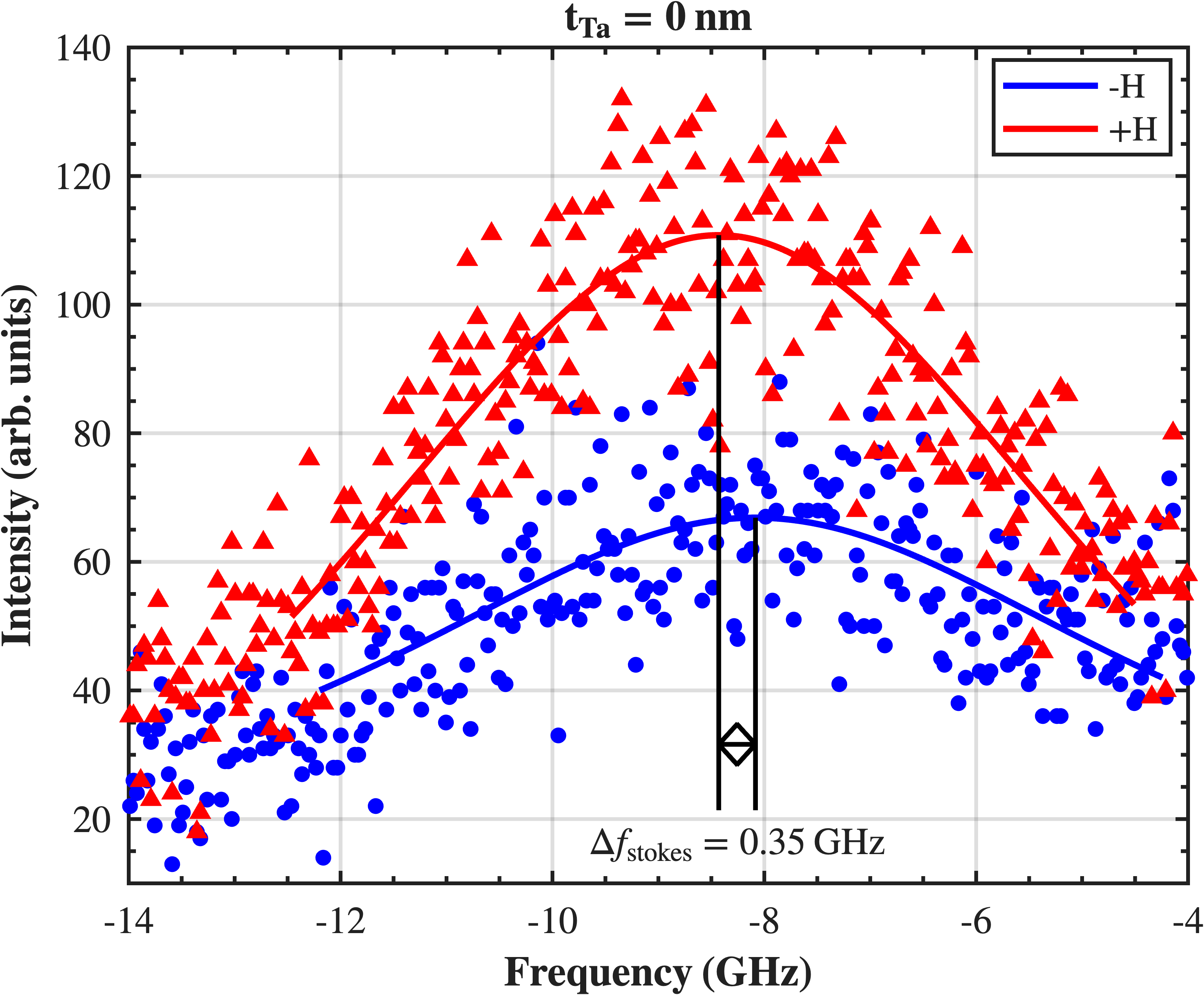}};
            \node[anchor=north west, xshift=-1.5mm, yshift=1mm, fill=white, fill opacity=0.8, text opacity=1, font=\tiny, inner sep=1.5pt] at (img.north west) {(a)};
        \end{tikzpicture}
    \end{minipage}\hfill
    \begin{minipage}[t]{0.48\linewidth}
        \centering
        \begin{tikzpicture}
            \node[inner sep=0pt] (img) {\includegraphics[width=\linewidth, height=0.12\textheight, keepaspectratio]{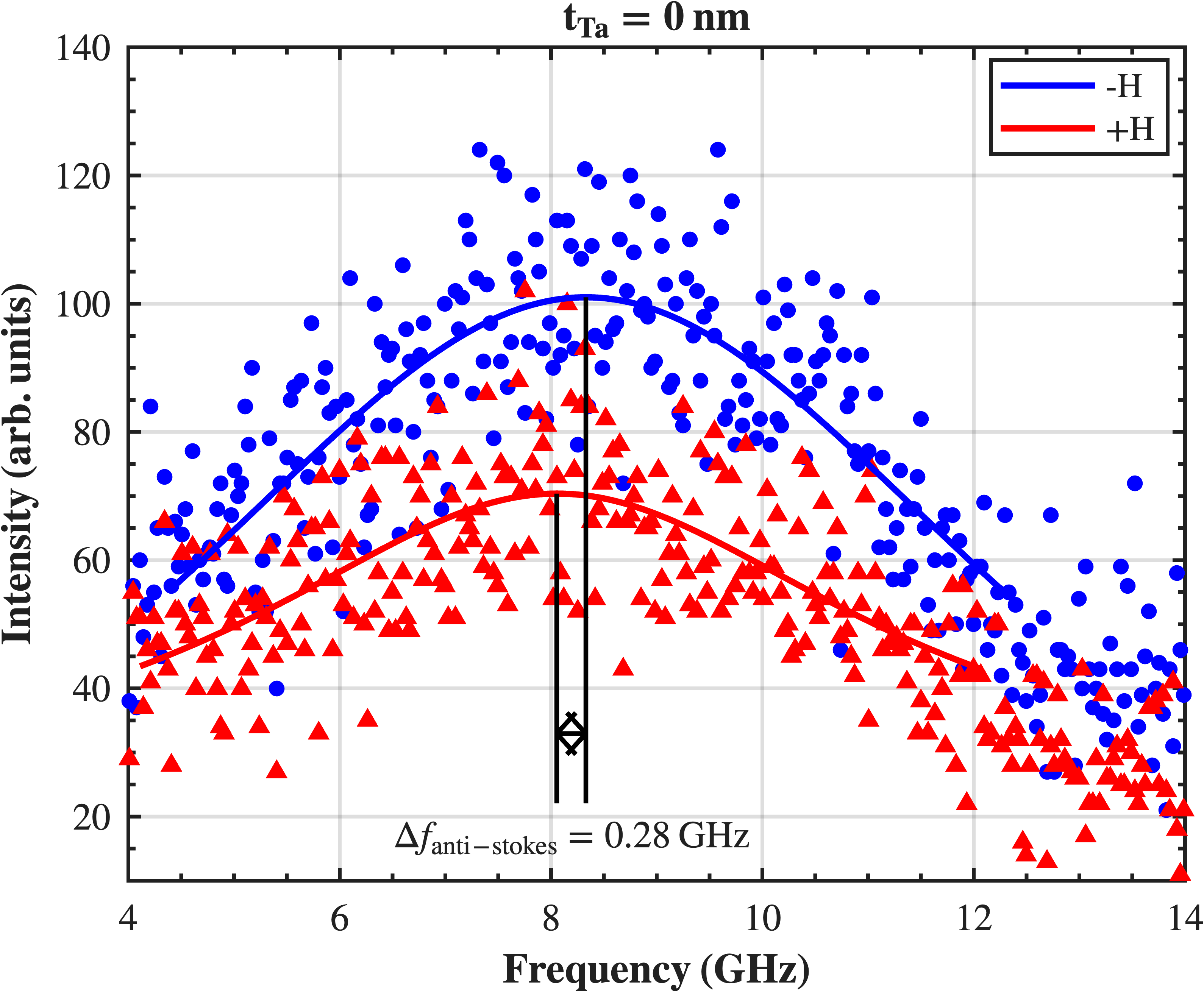}};
            \node[anchor=north west, xshift=-1.5mm, yshift=1mm, fill=white, fill opacity=0.8, text opacity=1, font=\tiny, inner sep=1.5pt] at (img.north west) {(b)};
        \end{tikzpicture}
    \end{minipage}

    \vspace{1mm}

    % --- ROW 2 (0.1 nm) ---
    \begin{minipage}[t]{0.48\linewidth}
        \centering
        \begin{tikzpicture}
            \node[inner sep=0pt] (img) {\includegraphics[width=\linewidth, height=0.12\textheight, keepaspectratio]{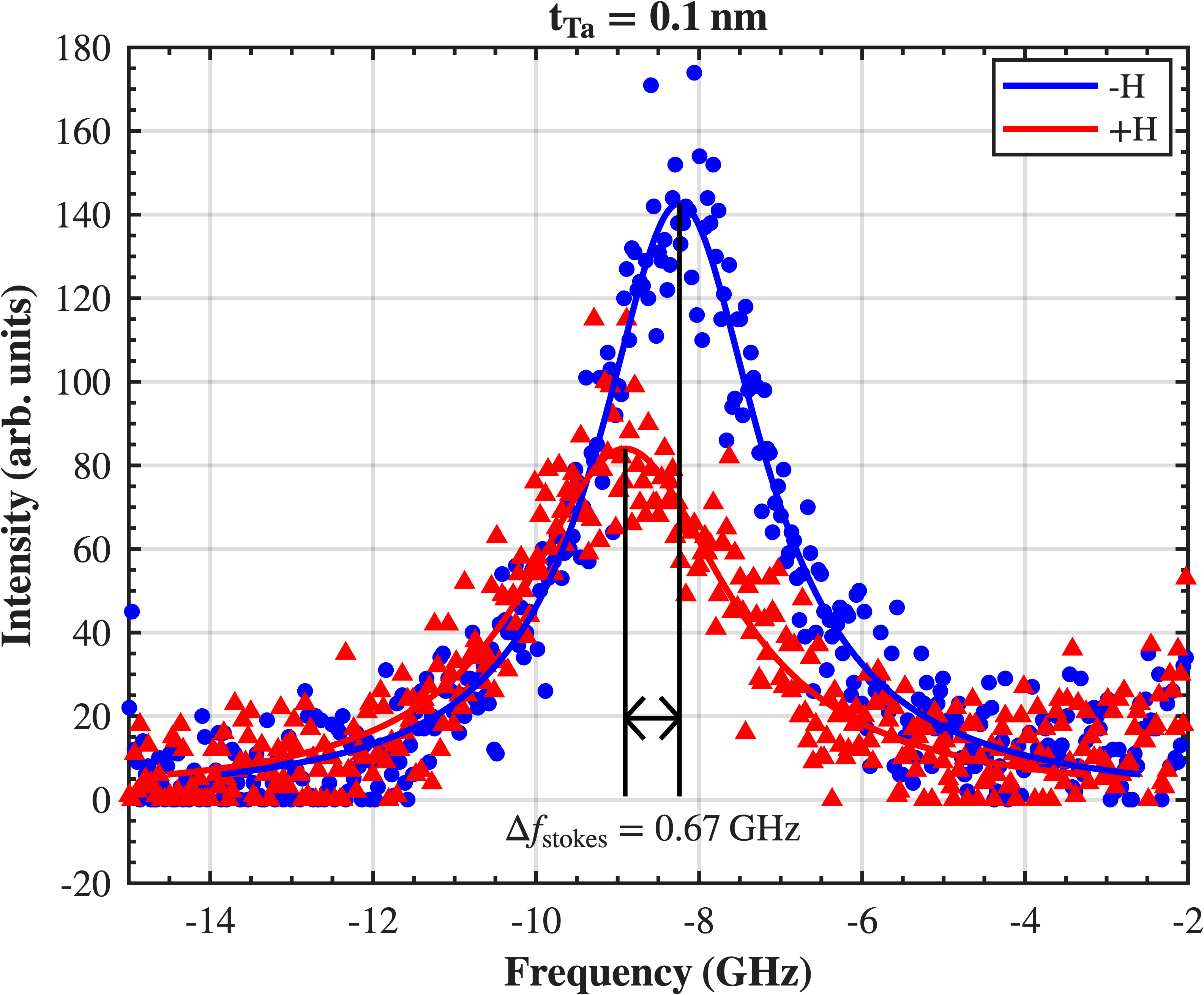}};
            \node[anchor=north west, xshift=-1.5mm, yshift=1mm, fill=white, fill opacity=0.8, text opacity=1, font=\tiny, inner sep=1.5pt] at (img.north west) {(c)};
        \end{tikzpicture}
    \end{minipage}\hfill
    \begin{minipage}[t]{0.48\linewidth}
        \centering
        \begin{tikzpicture}
            \node[inner sep=0pt] (img) {\includegraphics[width=\linewidth, height=0.12\textheight, keepaspectratio]{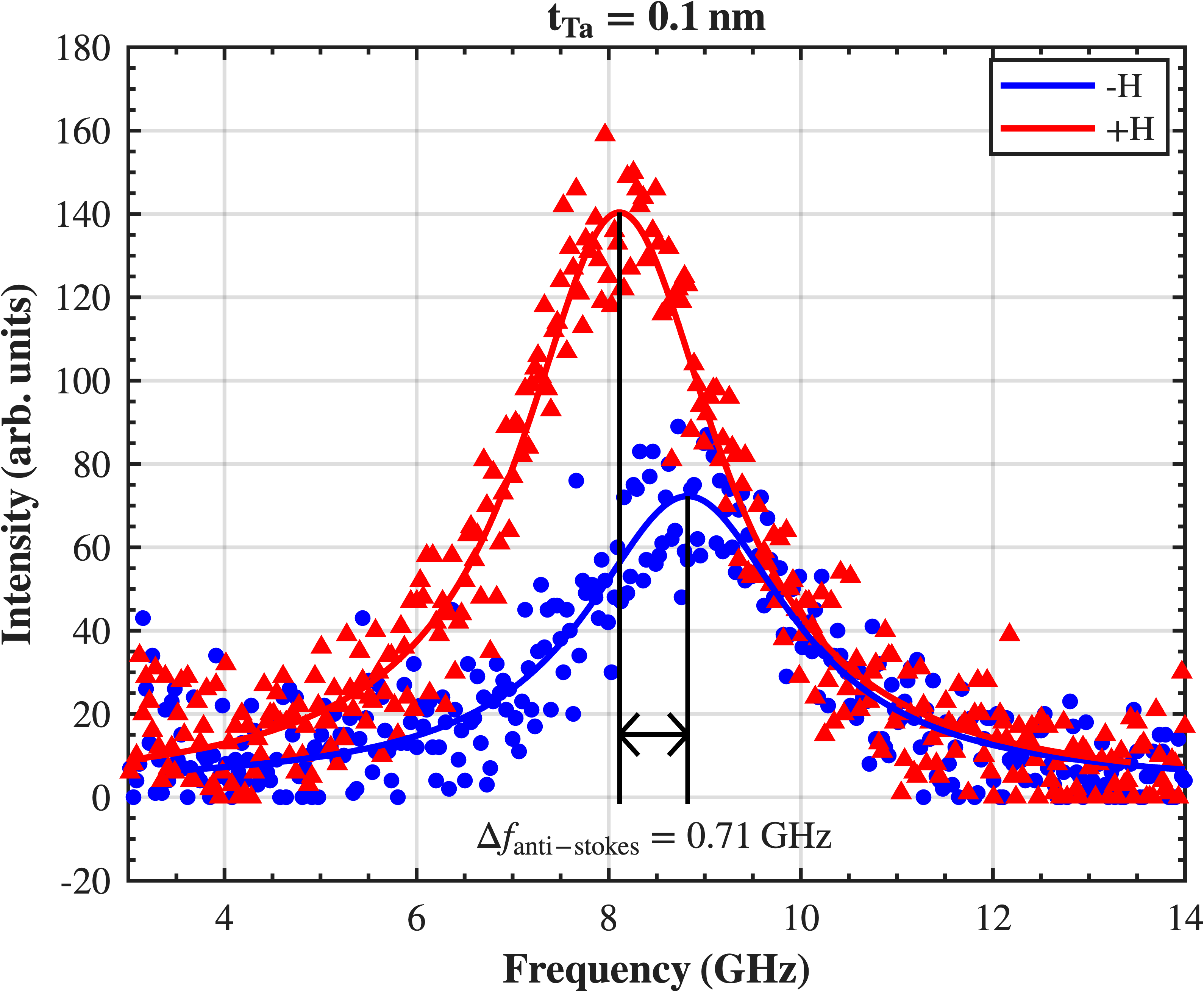}};
            \node[anchor=north west, xshift=-1.5mm, yshift=1mm, fill=white, fill opacity=0.8, text opacity=1, font=\tiny, inner sep=1.5pt] at (img.north west) {(d)};
        \end{tikzpicture}
    \end{minipage}

    \vspace{1mm}

    % --- ROW 3 (0.2 nm) ---
    \begin{minipage}[t]{0.48\linewidth}
        \centering
        \begin{tikzpicture}
            \node[inner sep=0pt] (img) {\includegraphics[width=\linewidth, height=0.12\textheight, keepaspectratio]{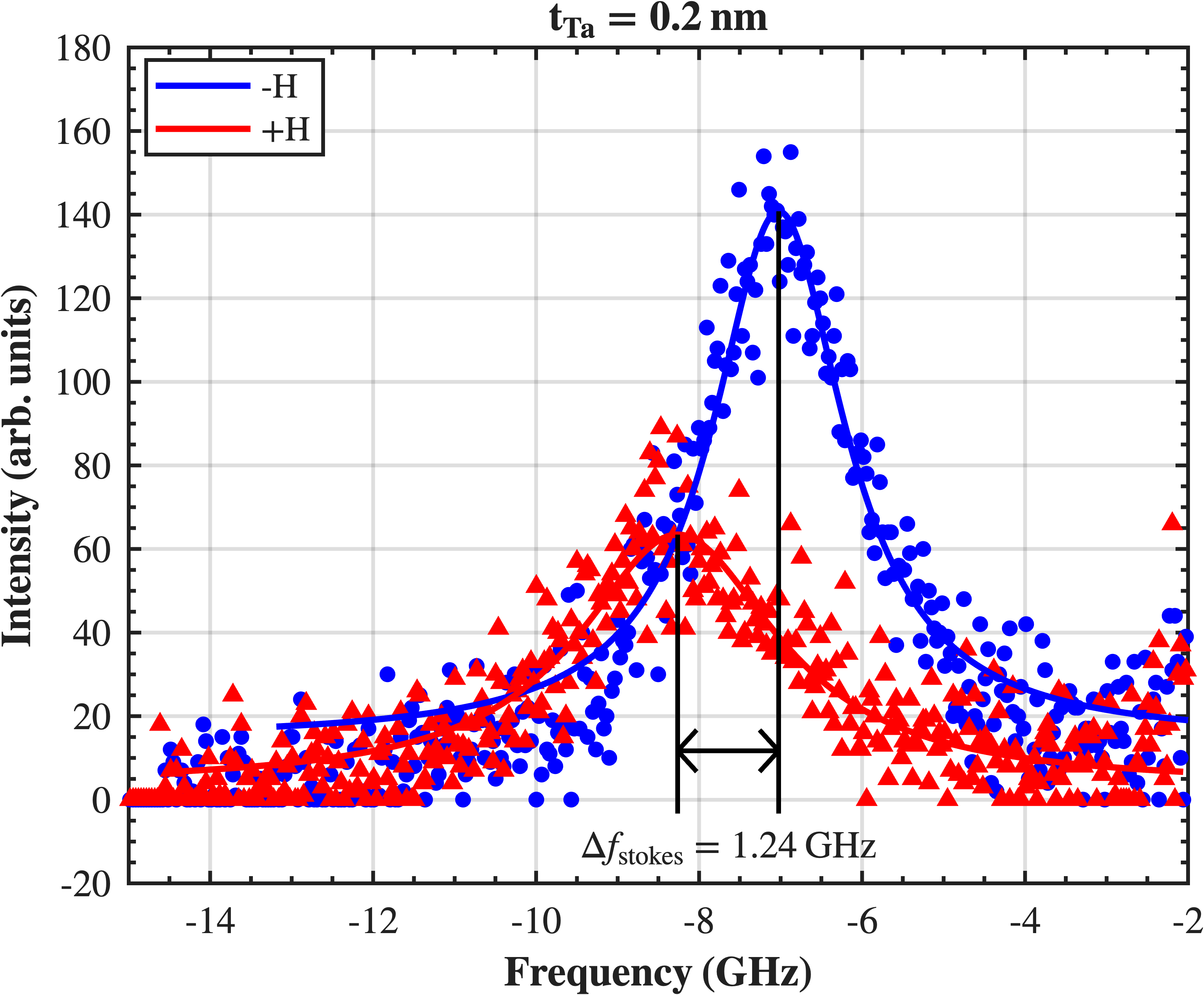}};
            \node[anchor=north west, xshift=-1.5mm, yshift=1mm, fill=white, fill opacity=0.8, text opacity=1, font=\tiny, inner sep=1.5pt] at (img.north west) {(e)};
        \end{tikzpicture}
    \end{minipage}\hfill
    \begin{minipage}[t]{0.48\linewidth}
        \centering
        \begin{tikzpicture}
            \node[inner sep=0pt] (img) {\includegraphics[width=\linewidth, height=0.12\textheight, keepaspectratio]{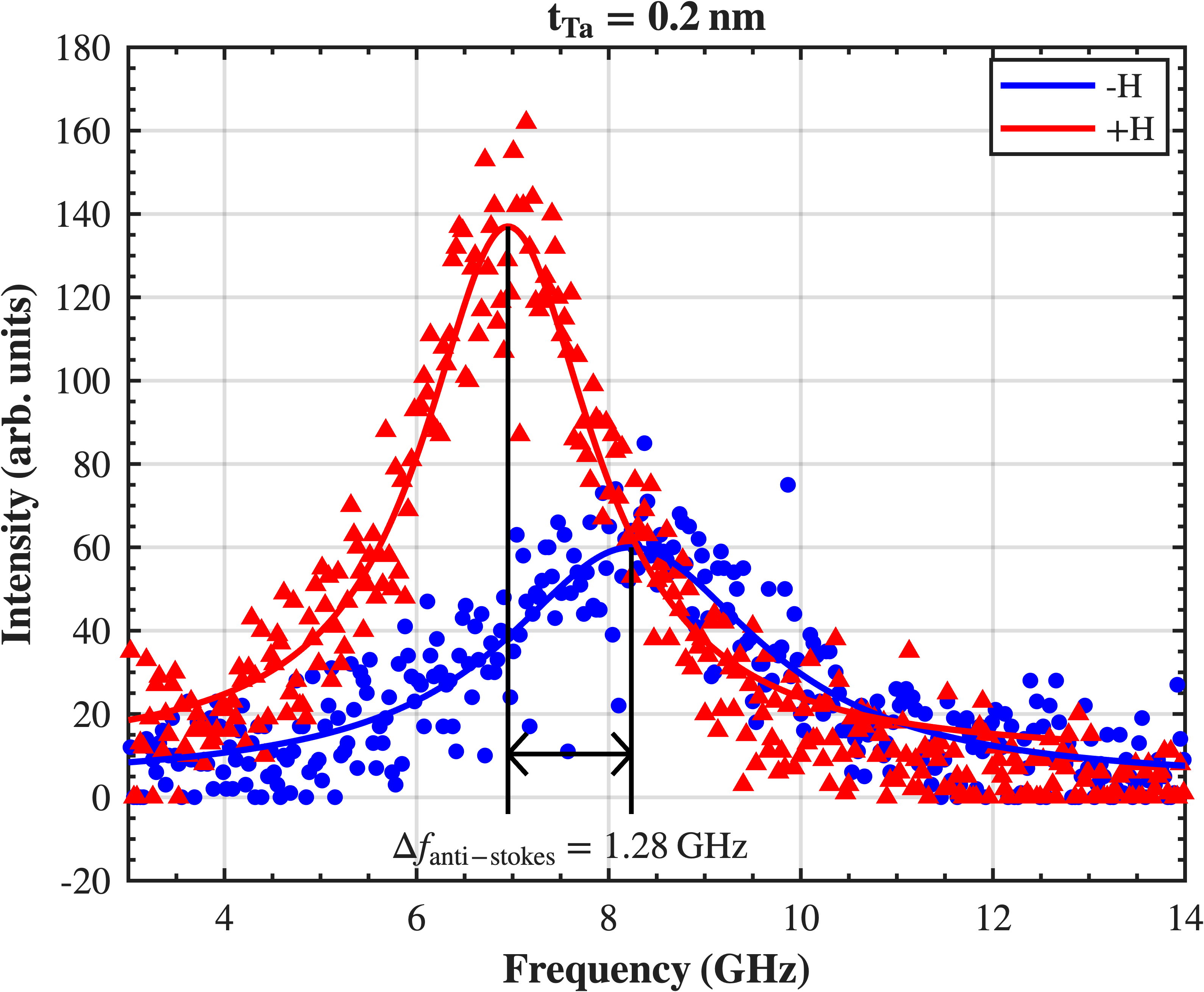}};
            \node[anchor=north west, xshift=-1.5mm, yshift=1mm, fill=white, fill opacity=0.8, text opacity=1, font=\tiny, inner sep=1.5pt] at (img.north west) {(f)};
        \end{tikzpicture}
    \end{minipage}

    \vspace{1mm}

    % --- ROW 4 (0.4 nm) ---
    \begin{minipage}[t]{0.48\linewidth}
        \centering
        \begin{tikzpicture}
            \node[inner sep=0pt] (img) {\includegraphics[width=\linewidth, height=0.12\textheight, keepaspectratio]{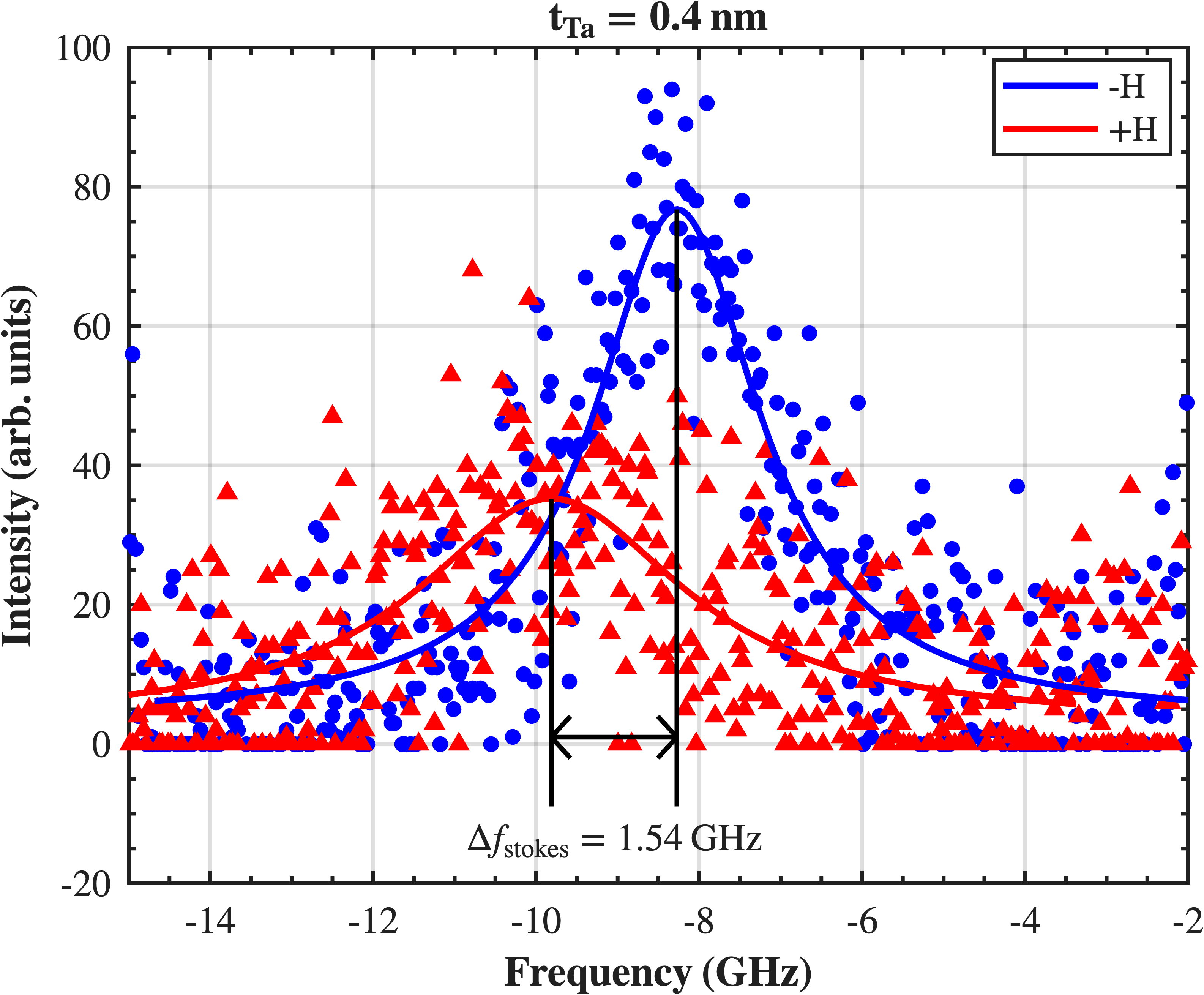}};
            \node[anchor=north west, xshift=-1.5mm, yshift=1mm, fill=white, fill opacity=0.8, text opacity=1, font=\tiny, inner sep=1.5pt] at (img.north west) {(g)};
        \end{tikzpicture}
    \end{minipage}\hfill
    \begin{minipage}[t]{0.48\linewidth}
        \centering
        \begin{tikzpicture}
            \node[inner sep=0pt] (img) {\includegraphics[width=\linewidth, height=0.12\textheight, keepaspectratio]{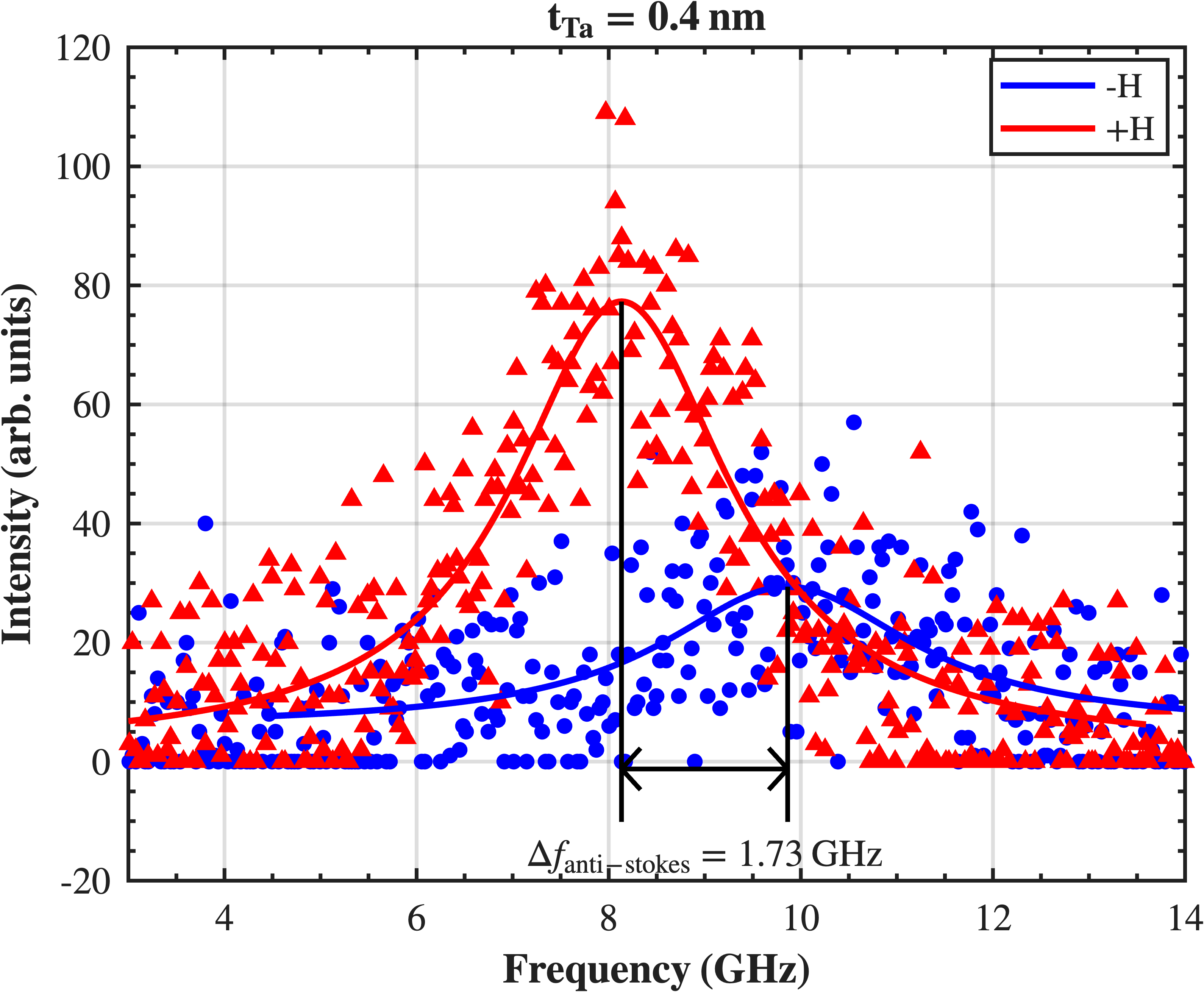}};
            \node[anchor=north west, xshift=-1.5mm, yshift=1mm, fill=white, fill opacity=0.8, text opacity=1, font=\tiny, inner sep=1.5pt] at (img.north west) {(h)};
        \end{tikzpicture}
    \end{minipage}

  \vspace{2mm}

    \caption{BLS spectra for samples with Ta layer thicknesses of (a)--(b) $t_{\mathrm{Ta}} = 0\,\mathrm{nm}$, (c)--(d) $t_{\mathrm{Ta}} = 0.1\,\mathrm{nm}$, (e)--(f) $t_{\mathrm{Ta}} = 0.2\,\mathrm{nm}$, and (g)--(h) $t_{\mathrm{Ta}} = 0.4\,\mathrm{nm}$ on Si substrates. Triangular (red) and circular (blue) symbols denote experimental data measured under positive ($+H$) and negative ($-H$) IP saturation magnetic fields, respectively. Solid lines represent Lorentzian fits to the measured spectra (blue for $+H$, red for $-H$). The nonreciprocal frequency shift is defined as $\Delta f_{\mathrm{S/AS}} = \left|f_{\mathrm{S/AS}}(+H) - f_{\mathrm{S/AS}}(-H)\right|$, where $f_{\mathrm{S/AS}}(\pm H)$ denotes the resonance peak frequency of the Stokes (S) or anti-Stokes (AS) mode.}
    \label{fig:bls}
\end{figure}

\pagebreak
\bibliography{DMI_Paper}

\end{document}